\documentclass[aip,amsmath,amssymb,reprint]{revtex4-1}
\usepackage[utf8]{inputenc}
\usepackage[T1]{fontenc}
\usepackage{graphicx}
\usepackage{xcolor}
\usepackage{times}
\usepackage{todonotes}
\usepackage{nicefrac}
\newcommand{\bd}[1]{#1}

\begin{document}
\preprint{AIP/123-QED}

\title{Peculiarities of \bd{escape} kinetics in the presence of \bd{athermal} noises}
\author{Karol Capa{\l}a}
\email{karol@th.if.uj.edu.pl} \affiliation{Marian Smoluchowski
Institute of Physics, and Mark Kac Center for Complex Systems
Research, Jagiellonian University, ul. St. {\L}ojasiewicza 11,
30--348 Krak\'ow, Poland}

\author{Bart{\l}omiej Dybiec}
\email{bartek@th.if.uj.edu.pl}  \affiliation{Marian Smoluchowski
Institute of Physics, and Mark Kac Center for Complex Systems
Research, Jagiellonian University, ul. St. {\L}ojasiewicza 11,
30--348 Krak\'ow, Poland}

\author{Ewa Gudowska-Nowak}
\email{ewa.gudowska-nowak@uj.edu.pl}  \affiliation{Marian Smoluchowski Institute of Physics, and Mark Kac Center for Complex Systems
Research, Jagiellonian University, ul. St. {\L}ojasiewicza 11,
30--348 Krak\'ow, Poland}

\date{\today}

\begin{abstract}
  Stochastic evolution of \bd{various dynamic systems and reaction networks is} commonly described in terms of noise assisted escape of an overdamped particle from a potential well, as devised by the paradigmatic Langevin equation \bd{in which additive Gaussian stochastic force reproduces effects of thermal fluctuations from the reservoir}. When implemented for systems close to equilibrium, the approach correctly explains emergence of Boltzmann distribution for the ensemble of trajectories generated by Langevin equation and relates intensity of the noise strength to the mobility.
  This scenario can be further generalized  to include effects of \bd{non-Gaussian}, burst-like forcing modeled by L\'evy noise. \bd{In this case however, the pulsatile additive noise cannot be treated as the internal (thermal), since the relation between the strength of the friction and variance of the noise is violated}.  
  Heavy tails of L\'evy noise distributions not only facilitate escape kinetics, but more importantly, change the escape protocol by altering final stationary state to a non-Boltzmann, non-equilibrium form. As a result,
  contrary to the kinetics induced by a Gaussian white noise, escape rates in environments with L\'evy noise are determined not by the barrier height, but instead, by the barrier width.
  We further discuss consequences of simultaneous action of thermal and L\'evy noises on statistics of passage times and population of reactants in double-well potentials. 
 
\end{abstract}

\pacs{
 05.40.Fb, 
 05.10.Gg, 
 02.50.-r, 
 02.50.Ey, 
 }

\maketitle

%
%
%

\textbf{Noise induced escape over a static potential barrier is the scenario underlying various fluctuations induced effects.
Numerous research explored Gaussian noise and L\'evy noise driven kinetics in double-well potential wells.
These two kinetics fundamentally differs, as they correspond to the continuous (Gaussian) and discontinuous (L\'evy) trajectories which in turn are responsible for very different escape protocols.
Here, we study the archetypal models of overdamped, stochastic dynamics in double-well potentials driven by a single L\'evy noise or a mixture of L\'evy and Gaussian noises.
Therefore, within the current studies,  we extend understanding of escape processes over a static potential barrier.
We explore the role of underlying assumptions by comparing results of numerical simulations with asymptotic scaling predicted by various approximations.
We show how the escape protocol is affected by abnormally long jumps (outliers) and what is the role of the central part of the jump length distribution.
We demonstrate that for the combined action of Gaussian and L\'evy noise sources various asymptotic regimes can be obtained.
}

\section{Introduction \label{sec:intro}}

Non-Gaussian L\'evy noises and L\'evy statistics are frequently objects of studies in the context of extreme, catastrophic events like economic crises \cite{stanley1986,mantegna2000}, outburst of epidemics \cite{newman1999} or millennial climate changes \cite{ditlevsen1999anomalous}. The increasing number of observations indicates presence of non-Gaussian fluctuations in the variety of complex dynamical systems
ranging from financial time series \cite{bouchaud1990} and recordings of turbulent behavior \cite{shlesinger1986b}, superdiffusion of micellar systems \cite{bouchaud1991} and transmission of light in polidispersive materials \cite{barthelemy2008},  to photon scattering in hot atomic vapors \cite{mercadier2009levyflights}, anomalous diffusion in laser cooling \cite{cohen1990,barkai2014}, 
 gaze dynamics \cite{amor2016} and memory retrieval in humans \cite{reynolds2007}. 
As a natural generalization of the Brownian motion, the L\'evy process is characterized by uncorrelated jumps sampled from the heavy-tailed, stable distribution of lengths and has been  extensively studied in a large number of theoretical and numerical considerations \cite{metzler2000,barkai2001,anh2003,brockmann2002,chechkin2006,jespersen1999,yanovsky2000,schertzer2001}.
Contrary to the Wiener process -- a mathematical abstract of the Brownian motion -- trajectories in the L\'evy motion are discontinuous, thus representative for pulsatile, irregular flickering. 
Whereas a prominent feature of the Brownian motion is a linear growth of the variance of the position with time -- this growth becomes faster (superlinear) for L\'evy motion.  Also, unlike  equilibrium noise which refers to the jump sizes distributed according to the Gaussian statistics of finite variance, its nonequilibrium counterpart,  the L\'evy (non-Gaussian) noise, describes the processes with large outliers and has diverging variance.


Importantly, L\'evy motions (called otherwise L\'evy flights (LF)) have been shown to  break detailed balance and microscopic reversibility \cite{garbaczewski2011levy,kusmierz2016breaking}.
Lack of detailed balance for the  Langevin dynamics with L\'evy flights  has important thermodynamic consequences: 
In static, periodic potentials with broken spatial symmetry solely action of the L\'evy noise induces the directed transport \cite{dybiec2008d}.
The key element of the ratcheting effect \cite{magnasco1993,reimann2002} is the acceleration of the escape process into the direction of the steeper slope of the potential.
This acceleration of the transport over the narrower potential barrier plays an important role in the L\'evy noise driven Kramers problem \cite{ditlevsen1999anomalous}.
In the weak noise limit, escape from the potential well induced by L\'evy noise is always faster \cite{imkeller2010hierarchy} than the analogous process induced by the Gaussian white noise and 
the most probable escape path is executed via a single long jump. This causes  the mean first passage time (MFPT) to depend dominantly on the barrier width $\delta$, i.e. $T \propto \delta^\alpha$ instead of barrier height $\Delta E$, i.e. $T\propto\exp(\beta \Delta E)$, typical for the Kramers kinetics in the presence of thermal (Gaussian) noise.
A similar, fully tractable analytically, solution  to the first passage time problem is observed for escapes from bounded domains under the action of L\'evy flights \cite{blumenthal1961, getoor1961, kac1950distribution, widom1961stable, kesten1961random}.

In line with Kramers approach the kinetic mechanism of a chemical reaction is described by means of a diffusion process along an internal coordinate $x$ whose stationary states before and after the reaction correspond to the minima of a double-well potential $V(x)$ located at $x_1$ and $x_2$ and separated by an energy barrier \cite{kramers1940}, see Fig.~\ref{fig:linearpotential}. Assuming local equilibrium in the internal space allows one to formulate the Gibbs equation and identify the diffusion currents in terms of kinetic equations balancing the reactant and product concentrations \cite{rubi}. Furthermore, derivation of rate constants for forward and reverse reactions  gives the ratio (the equilibrium constant)
\begin{eqnarray}
k_+/k_-=\exp[\beta (V(x_1)-V(x_2))]
    \end{eqnarray}
which in the ideal case depends only on \bd{system's} temperature via the Boltzmann coefficient $\beta$.

Within the paper we discuss escape from the potential wells induced/affected by L\'evy noises and analyze departure from equilibrium kinetics as expressed by the above equilibrium constant. Asymptotic properties of  systems driven by L\'evy noise  can be studied by the known L\'evy-It\^o decomposition \cite{imkeller2006,imkeller2006b} of L\'evy flights in terms of the sum of a Poisson compound process and a Gaussian white noise. Consequently, anomalous long jumps, which determine escape kinetics over the barriers, are represented by the Poissonian component of the noise. 

The paper starts with an introduction of a generic model system described by a Langevin equation (Section~\ref{sec:models}), followed by presentation of simulations' details.  Results  derived for various double-well potentials are discussed in Section~\ref{sec:results}. In the same Section asymptotic properties of escape kinetics affected by the combined action of the L\'evy and Gaussian noises are analyzed. The paper concludes with a summary (Section~\ref{sec:summary}) referring to main results and potential research areas.

%
%
%

\section{Model \label{sec:models}}
The barrier-crossing  is modeled in terms of spatially diffusive (overdamped) motion of a particle subject to the deterministic force field $f(x)=-V'(x)$ and a fluctuating force $\xi(t)$ describing interactions of the system with its environment:
\begin{equation}
\frac{dx}{dt}=-\frac{dV(x)}{dx} + \sigma \xi (t).
\label{eq:langevin}
\end{equation}
Here $\sigma$ is a parameter measuring intensity of fluctuations which equals $\sigma=\sqrt{2/\beta}$ for the motion of a Brownian particle in the strong friction limit.

We further assume that fluctuating force $\xi(t)$ is not Gaussian but instead, can be represented
as a formal time derivative of the symmetric $\alpha$-stable motion \cite{janicki1994b} $L(t)$, whose
characteristic function $\phi(k)=\langle \exp[ikL(t)] \rangle$ attains the form
\begin{equation}
 \phi(k)=\exp\left( - t \sigma^\alpha |k|^\alpha \right).
 \label{eq:fcharakt}
\end{equation}

The stochastic process $\{X(t), t\geqslant 0\}$ described by Eq.~(\ref{eq:langevin}) has increments
\begin{eqnarray}
\label{eq:discretization}
 \Delta x & = & x(t+\Delta t)-x(t) \\
 & = & -V'(x(t))\Delta t +\Delta t^{1/\alpha} \sigma \xi_t, \nonumber
\end{eqnarray}
where $\xi_t$ represents a sequence of independent, identically distributed random variables  \cite{chambers1976,weron1995,weron1996} following the symmetric $\alpha$-stable density \cite{janicki1994,janicki1996} with the unity scale parameter.
\bd{The scale parameter $\sigma$ becomes an independent, external parameter.}
For a clarity of the presentation, the scale parameter $\sigma$ in Eqs.~(\ref{eq:langevin}) and (\ref{eq:discretization}) is extracted from the noise definition, see Eq.~(\ref{eq:fcharakt}).

\begin{figure}[h!]
\centering
\includegraphics[width=0.9\columnwidth]{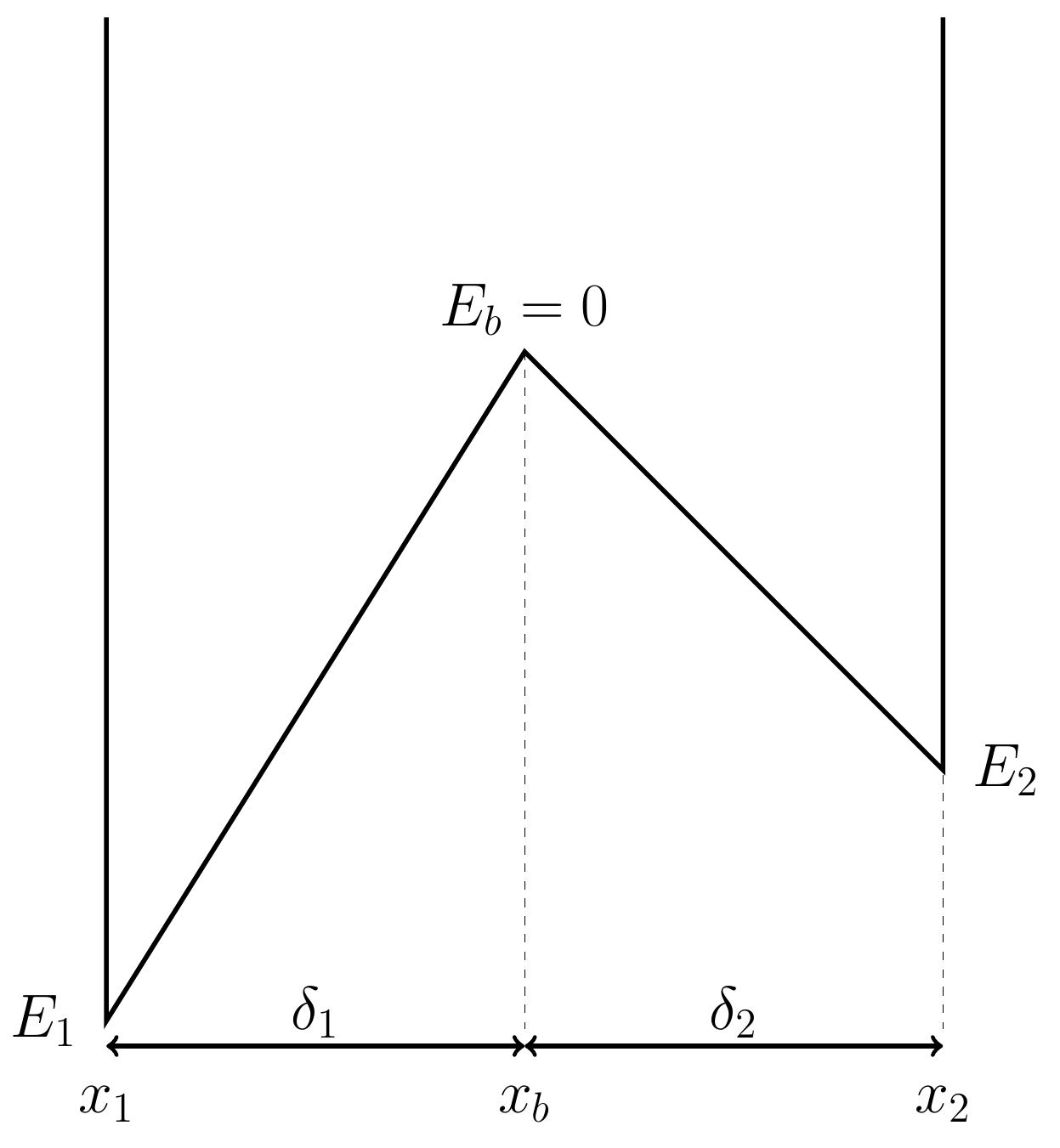}\\
\includegraphics[width=0.45\textwidth]{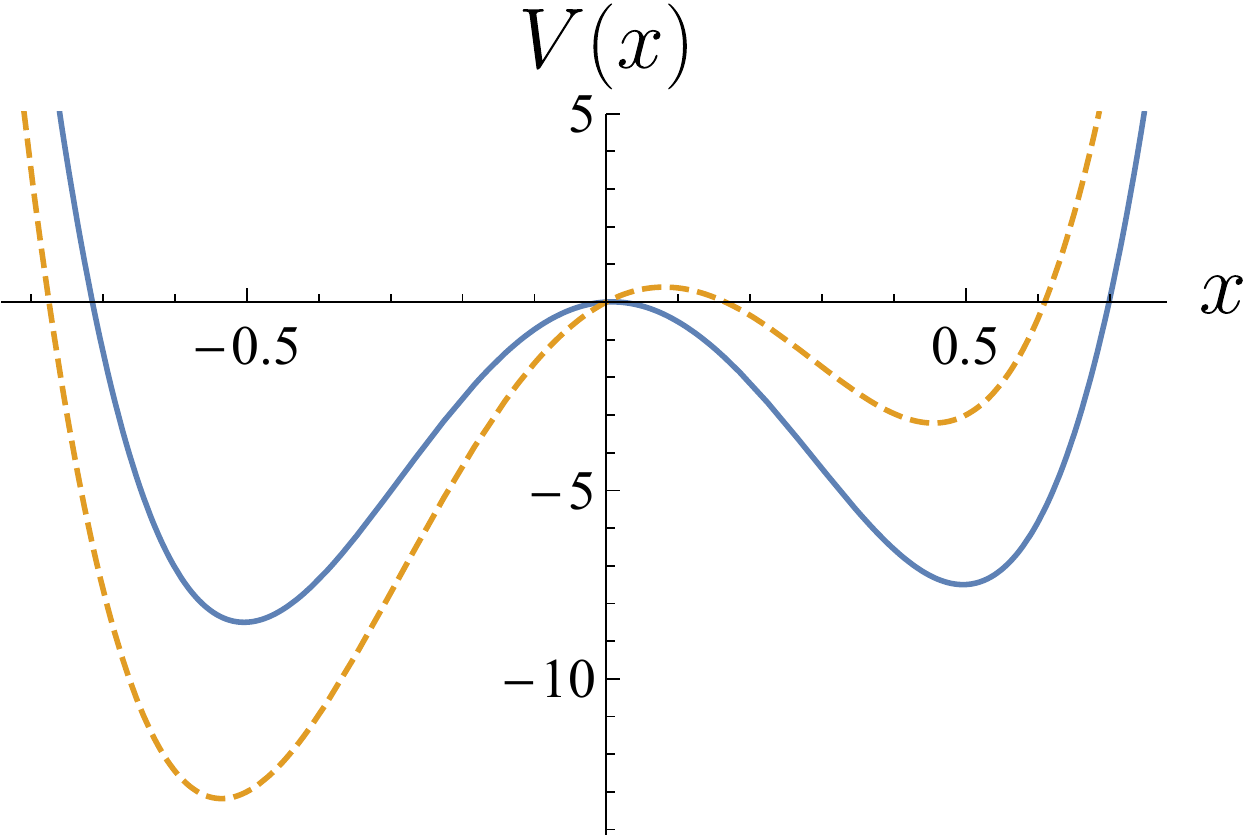}\\
\caption{Piecewise-linear, double-well potential (top panel) and the continuous double-well potential (bottom panel) used in the study.
The continuous potential $V(x) = 128 x^4-64 x^2 + a x$,
is given by Eq.~(\ref{eq:ciagly}) with the parameter $a$ controlling the potential asymmetry. Here, the solid line corresponds to $a=1$ and the dashed line to $a=10$.
}
\label{fig:linearpotential}
\end{figure}

Main properties of the escape kinetics can be drawn from the analysis of Eq.~(\ref{eq:discretization}):
For a motion in a piecewise-linear potential starting in the left potential minimum, see Fig.~\ref{fig:linearpotential}, the Euler approximation (\ref{eq:discretization}) reduces to
\begin{equation}
\Delta x=  - \frac{\Delta E_1}{\delta _1}\Delta t +\Delta t^{1/\alpha}  \sigma \xi_t  ,
\label{eq:linearforce}
\end{equation}
where $\Delta E_1 = E_b-E_1$ is the depth of the left potential well.
Without loss of generality, we can assume that  $E_b=0$, see top panel of Fig.~\ref{fig:linearpotential}.
The transition between potential wells includes the surmounting of the potential barrier, while the sliding along the potential slope is expected to be instantaneous.
Accordingly, the transition from the left to the right minimum of the potential is recorded for trajectories for which
\begin{equation}
\Delta x \geqslant \delta_1,
\label{eq:trcond}
\end{equation}
where $\delta_1$ is the distance from the left minimum of the potential to the barrier top.
From the discretization scheme~(\ref{eq:discretization}) and Eq.~(\ref{eq:trcond}) one gets the following condition
\begin{equation}
- \frac{\Delta E_1}{\delta _1}\Delta t + \Delta t^{1/\alpha} \sigma \xi   \geqslant \delta_1,
\label{eq:discretization-app}
\end{equation}
which results in
\begin{equation}
\xi \geqslant \xi_{1} = \frac{1}{\sigma \Delta t^{1/\alpha-1}}  \left[  \frac{\delta_1}{\Delta t} +  \frac{\Delta E_1}{\delta_1} \right]  .
\label{eq:jumplength}
\end{equation}
For the symmetric $\alpha$-stable density, the probability of observing a jump larger than $\xi_1$ is
\begin{equation}
P(\xi \geqslant \xi_1) \propto \xi_1^{-\alpha}.    
\end{equation}
Consequently, from Eq.~(\ref{eq:jumplength}) one obtains
\begin{equation}
P\left(\xi \geqslant \xi_1 \right) \propto \left(\frac{\delta_1}{\Delta t} + \frac{\Delta E_1}{\delta_1}\right) ^{-\alpha}.
\label{eq:k12}
\end{equation}
Analogously, for backward passages 
\begin{equation}
P\left(\xi \geqslant \xi_2 \right) \propto \left(\frac{\delta_2}{\Delta t} + \frac{\Delta E_2}{\delta_2}\right) ^{-\alpha}.
\label{eq:k21}
\end{equation}
Eqs.~(\ref{eq:k12}) and~(\ref{eq:k21}) define escape (transition) rates $k_{12}$, $k_{21}$ from the left/right potential wells:
\begin{equation}
\left(\frac{\delta_2+ \frac{\Delta E_2}{\delta_2} \Delta t}{\delta_1 + \frac{\Delta E_1}{\delta_1}\Delta t}\right)^{\alpha}\propto \frac{k_{12}}{k_{21}}
\label{eq:full}
    \end{equation}
For a typical chemical reaction scheme between reactants ($R$) and products ($P$), $R\rightleftharpoons P$, the ratio $k_{12}/k_{21}$ can be related at equilibrium to the mass action law and the equilibrium concentration of species \cite{gillespie1996mathematics}
\begin{equation}
\frac{k_{12}}{k_{21}}= \frac{P_2}{P_1}. 
\label{eq:population}
\end{equation}
Here $P_1$ and $P_2$ are (equilibrium, steady state) probabilities of finding the system in either (left/right) potential wells.
The probability $P_1(t)$ that the system is in the left state is given by
\begin{equation}
    P_1(t)=\int_{-\infty}^{x_b}p(x,t)dx,
    \label{eq:p1}
\end{equation}
where $p(x,t)$ is a time dependent probability density of finding a particle at time $t$ in the vicinity of $x$, and $x_b$ is the point separating left and right states. Analogously, the formula for $P_2(t)$ reads
\begin{equation}
    P_2(t)=\int_{x_b}^{\infty}p(x,t)dx=1-P_1(t).
    \label{eq:p2}
\end{equation}
If stationary $P_1$ and $P_2$ exist, they are obtained from the above integrals in the $t\to\infty$ limit with $p(x,t)$ replaced by the stationary density $p(x)$.

For a fixed potential barrier  Eq.~(\ref{eq:full}) reduces, in the $\Delta t \to 0 $ limit, to the situation considered in \cite{ditlevsen1999anomalous,imkeller2006,imkeller2006b}
\begin{equation}
\frac{P_2}{P_1}=\frac{k_{12}}{k_{21}} \propto  \left(\frac{\delta_2}{\delta_1}\right) ^{\alpha}.
\label{eq:ditlevsen}
\end{equation}
At the same time, for fixed $\Delta t$ and a high barrier ($\Delta E \gg 1/\Delta t$) one may obtain \cite{bier2018}
\begin{equation}
\frac{P_2}{P_1}=\frac{k_{12}}{k_{21}} \propto \left(\frac{ \Delta E_2}{\Delta E_1}\right)^{\alpha}.
\label{eq:bier}
\end{equation}
The scalings predicted by Eqs.~(\ref{eq:ditlevsen}) and~(\ref{eq:bier}) should be contrasted with the Gaussian white noise limit, in which the  ratio of Kramers rates  \cite{kramers1940,hanggi1990} leads to
\begin{equation}
    \frac{P_2}{P_1}=\frac{k_{12}}{k_{21}} \propto \exp\left[ \frac{E_2-E_1}{\sigma^2}  \right].
    \label{eq:gauss}
\end{equation}

Within the stochastic description of chemical kinetics, the transition rates can be conveniently defined in terms of inverse of the mean first passage time (MFPT), e.g. $k_{12}=T_{12}^{-1}$
where 
\begin{eqnarray}
    T_{12} & = & \langle \tau \rangle \\
    &=& \langle \min\{\tau : x(0)=x_1=-\delta_1 \;\mbox{and}\; x(\tau) \geqslant x_b  \} \rangle. \nonumber
    \label{eq:mfpt-def}
\end{eqnarray}
For Gaussian noise ($\alpha=2$) entering Eq.~(\ref{eq:langevin}) the MFPT can be calculated exactly \cite{gardiner2009} and reads
\begin{eqnarray}
\label{eq:mfpt}
    T(x_0 \to x_b) & = &  \frac{1}{\sigma^2}\int_{x_0}^{x_b} dz \exp\left[ V(z)/\sigma^2 \right]  \\ \nonumber
    && \times \int_{-\infty}^z dy \exp\left[ -V(y)/\sigma^2  \right],
\end{eqnarray}
while for $\alpha<2$ one needs to rely either on stochastic simulations and scaling analysis \cite{chechkin2007,chechkin2005} or on a numerical solution of the corresponding fractional diffusion equation. 

For the purpose of further analysis we define quotients $\mathcal{P}=P_2/P_1$ and $\mathcal{T}=T(x_1)/T(x_2)=k_{12}/k_{21}$, where in the last expression indices refer to the location  of the left/right minimum of the potential. In order to consider L\'evy fluctuations embedded in confining (steep) potentials securing existence of variances of stationary states, we analyze motion in a piecewise-linear (cf. Fig.~\ref{fig:linearpotential}) and in a
continuous double-well potential
\begin{equation}
V(x) = 128 x^4-64 x^2 + a x.
\label{eq:ciagly}
\end{equation}
  In the latter form of $V(x)$ the parameter $a$ controls the potential asymmetry, depths of minima and their location. \bd{The coefficients of the polynomial terms have been chosen to secure that recrossing events are rare, for which the potential wells have to be deep enough and the barrier region sufficiently narrow \cite{dybiec2007}.
  Otherwise, especially for Gaussian white noise, discrimination between states is less apparent.
  } 
It should be stressed that both forms of potentials are sufficient to restrain the trajectories of L\'evy flights from infinite escapes
 \cite{chechkin2002,chechkin2003,chechkin2004,dybiec2010d} by introducing impermeable boundaries and (or) deep wells confining the motion.

%
%
%

\section{Results \label{sec:results}}

Results included in following subsections have been constructed numerically by methods of stochastic dynamics.
Eq.~(\ref{eq:langevin}) was integrated by the Euler-Maryuama method, see Eq.~(\ref{eq:discretization}), with the time step of integration $\Delta t=10^{-5}$ and averaged over  $10^4$ -- $10^5$ repetitions.  
We start with the study of properties of anomalous kinetics in piecewise-linear and continuous   potentials driven by a single L\'evy noise only (Sec.~\ref{sec:plp}).
Next, we focus on the combined action of Gaussian white noise and L\'evy noise (Sec.~\ref{sec:mixture}).
Finally, in order to further explore role of combined action of two noise sources we confront results of L\'evy noise-driven kinetics  with the problem of escape from a finite interval (Sec.~\ref{sec:finite}).


\subsection{Escape induced by L\'evy noise \label{sec:plp}}

Figure~\ref{fig:linearpotential} presents a sample  piecewise-linear (top panel) and a continuous (bottom panel) double-well potentials.
For convenience we choose a potential with maximum at $x_b=0$ and the maximal value $E_b=0$.
For such a potential we can easily control the ratio of widths $\delta_2/\delta_1$ and depths of potential wells.
The potential depicted in the top panel of Fig.~\ref{fig:linearpotential} gives the full flexibility and allows verification of various hypothesis regarding stochastic dynamics.
The continuous double-well potentials given by Eq.~(\ref{eq:ciagly}) with $a=1$ (blue solid line) and $a=10$ (orange dashed line) are depicted in the bottom panel of Fig.~\ref{fig:linearpotential}. 
For $a=1$, the depths of potential wells are $\Delta E_1\approx 8.5$, $\Delta E_2 \approx 7.5$ and the ratio of locations of minima $\delta_2/\delta_1\approx 0.98 $.
Analogously, for $a=10$, we have $\Delta E_1\approx 13.2$, $\Delta E_2 \approx 3.2$ and $\delta_2/\delta_1\approx 0.85$.
Piecewise-linear and continuous setups differs mainly with respect to the relative depth of potential wells and shape of the potential for $x<x_1$ and $x>x_2$ due to the way of restricting the domain of motion, compare top versus bottom panel of Fig.~\ref{fig:linearpotential} and Eq.~(\ref{eq:ciagly}).

\begin{figure}[h!]
\centering
\includegraphics[width=0.45\textwidth]{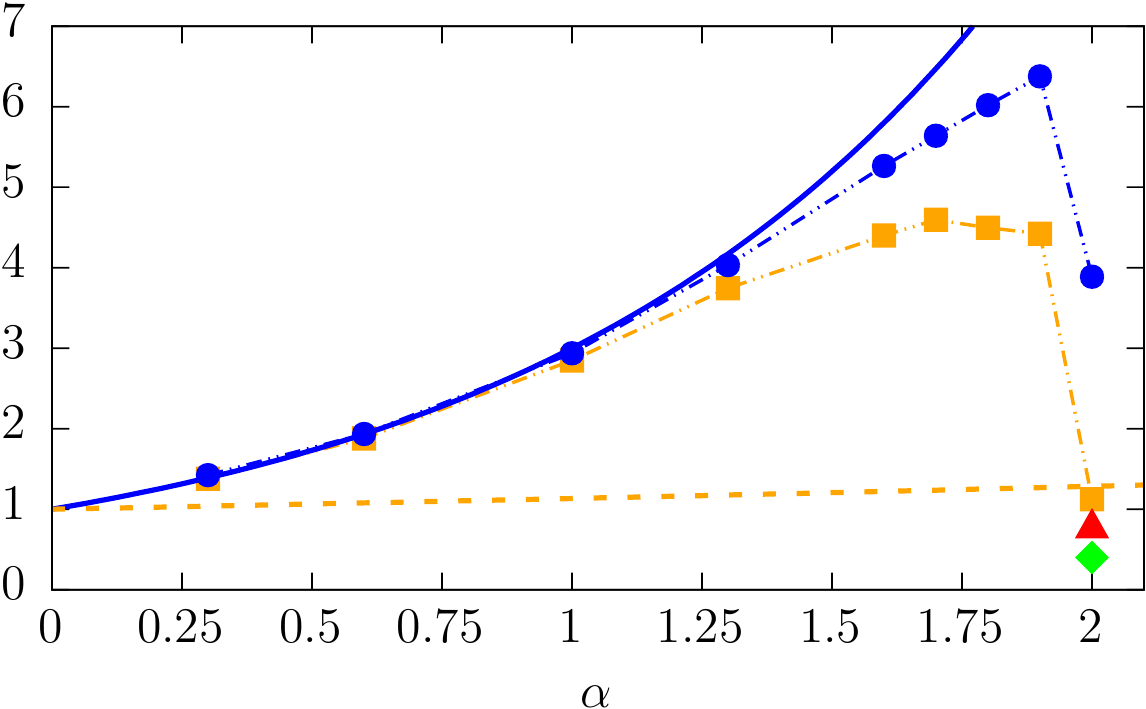} \\
\includegraphics[width=0.45\textwidth]{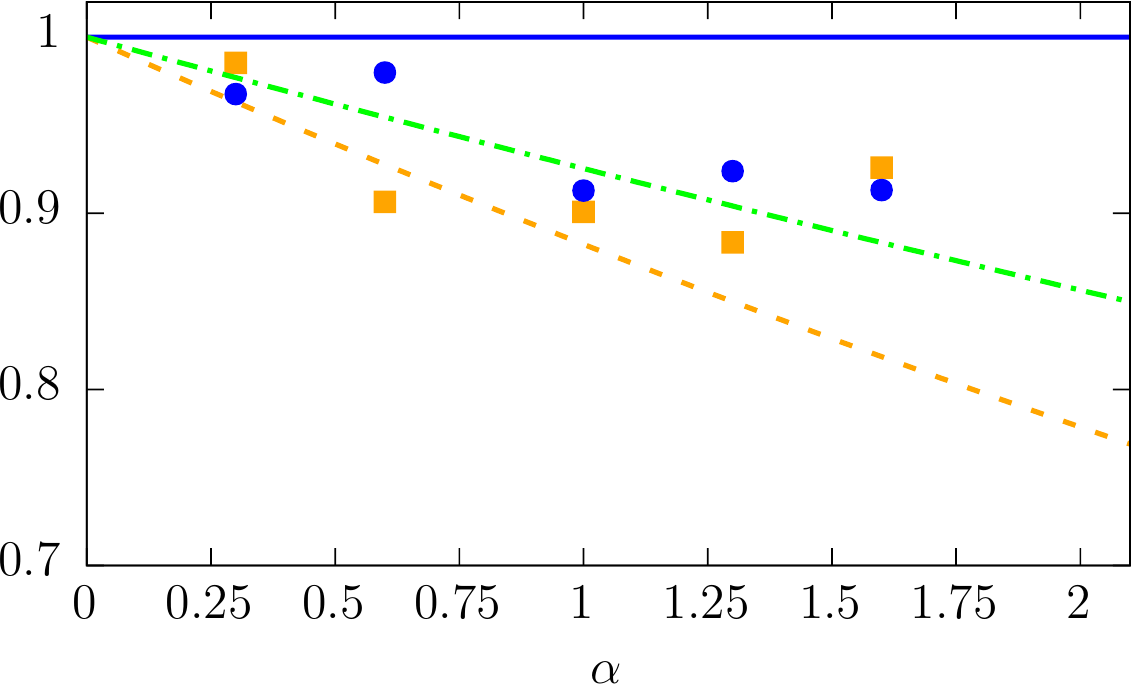} \\
\includegraphics[width=0.45\textwidth]{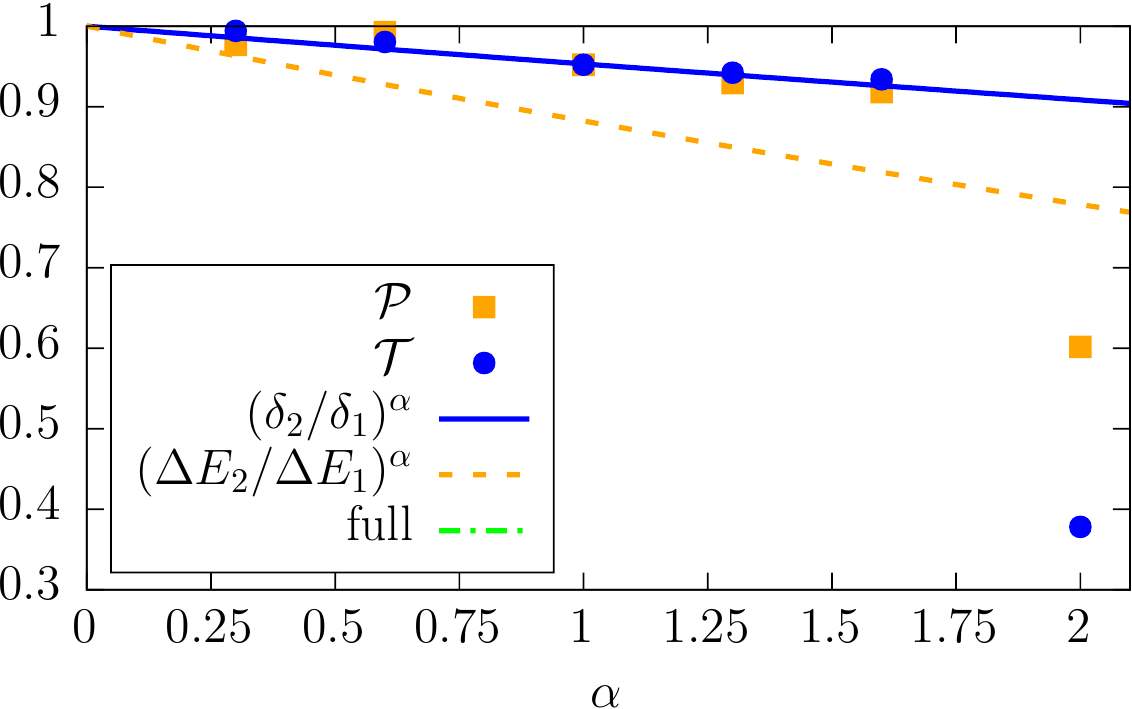} \\
\caption{
Symbols represent the ratios $\mathcal{P}$ of occupation probabilities (\textcolor{orange}{$\blacksquare$}) and $\mathcal{T}$ of transition rates  (\textcolor{blue}{$\bullet$}).
Results of simulations are displayed with points while lines show various theoretical scalings discussed in the text:  ``full'' (green dot-dashed, see Eq.~(\ref{eq:full})), ``width ratio'' (blue solid, see Eq.~(\ref{eq:ditlevsen})) and ``depth ratio'' (orange dashed, see Eq.~(\ref{eq:bier})).
Subsequent panels correspond to various setups:  piecewise-linear potential with $\Delta E_1=8.5$, $\Delta E_2=7.5$, $x_1=0.25$ and $x_2=0.75$ (top panel),
piecewise-linear potential with $\Delta E_1=85000$, $\Delta E_2=75000$, $x_1=0.7$ and $x_2=0.7$ (middle) and 
the continuous potential (\ref{eq:ciagly}) with $a=1$ (bottom).
The red triangle (\textcolor{red}{$\blacktriangle$}) and  the green rhombus (\textcolor{green}{$\blacklozenge$}) in the top panel depict analytical evaluation of $\mathcal{P}$, $\mathcal{T}$, respectively, derived with the stationary $p(x)$ for the Gaussian ($\alpha=2$) noise.
}
\label{fig:zat_E85}
\end{figure}

Results of numerical simulations with various parameters characterizing the piecewise-linear double-well potential, see Fig.~\ref{fig:linearpotential},  are depicted in top and middle panels of Fig.~\ref{fig:zat_E85}.
These results are compared and confronted with appropriate asymptotic formulas, see Eqs.~(\ref{eq:ditlevsen}), (\ref{eq:bier}) and the full formula (\ref{eq:full}).
Finally, findings for the piecewise-linear potential are also confronted with results for the continuous potential with $a=1$, see bottom panel of Fig~\ref{fig:zat_E85}.

\begin{figure}[h!]
\centering
\includegraphics[width=0.45\textwidth]{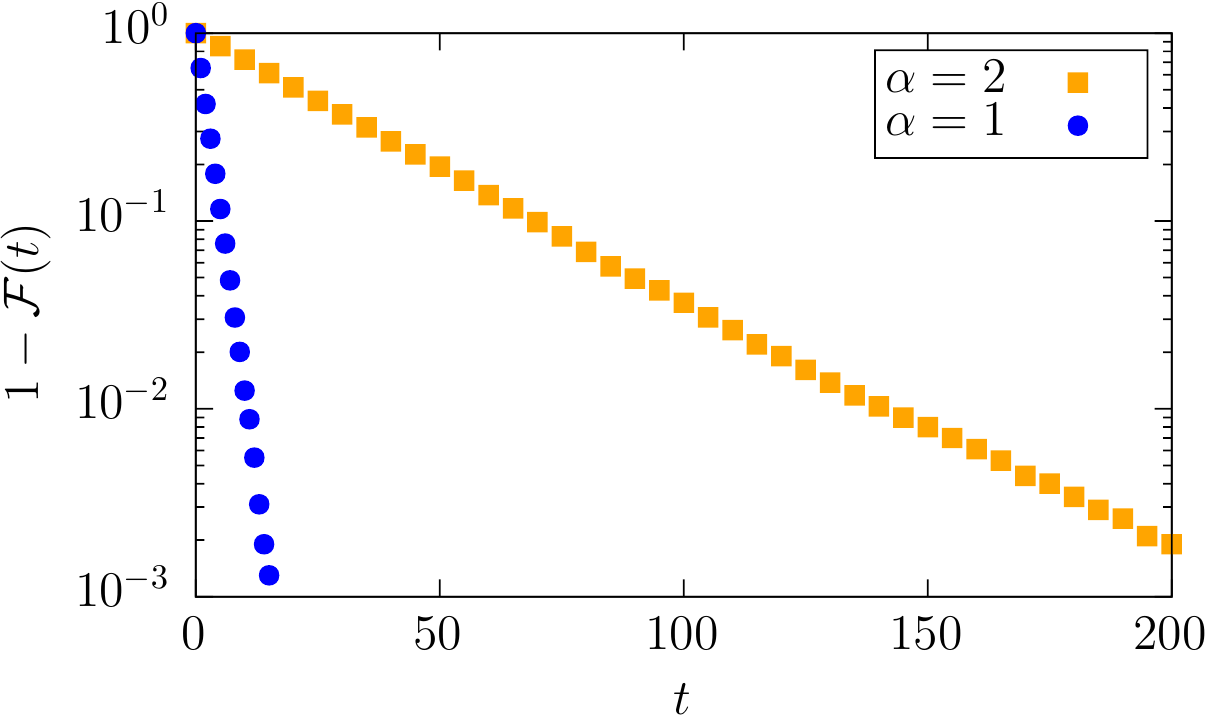}\\
\includegraphics[width=0.45\textwidth]{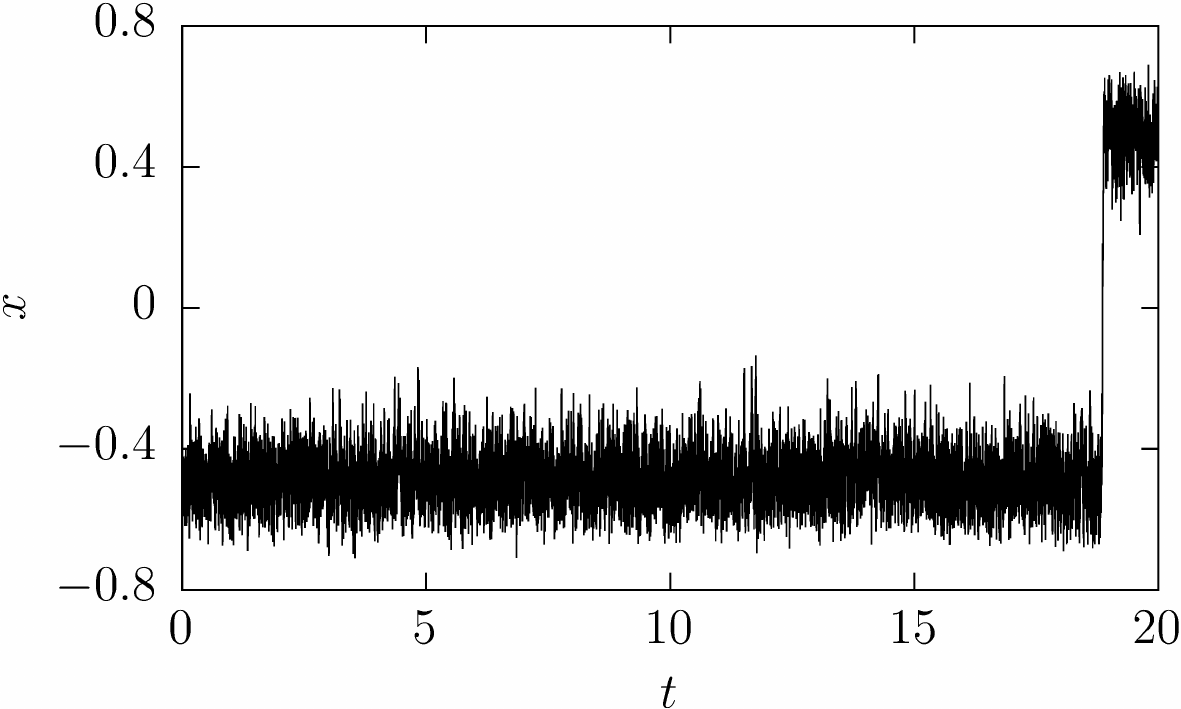}\\
\includegraphics[width=0.45\textwidth]{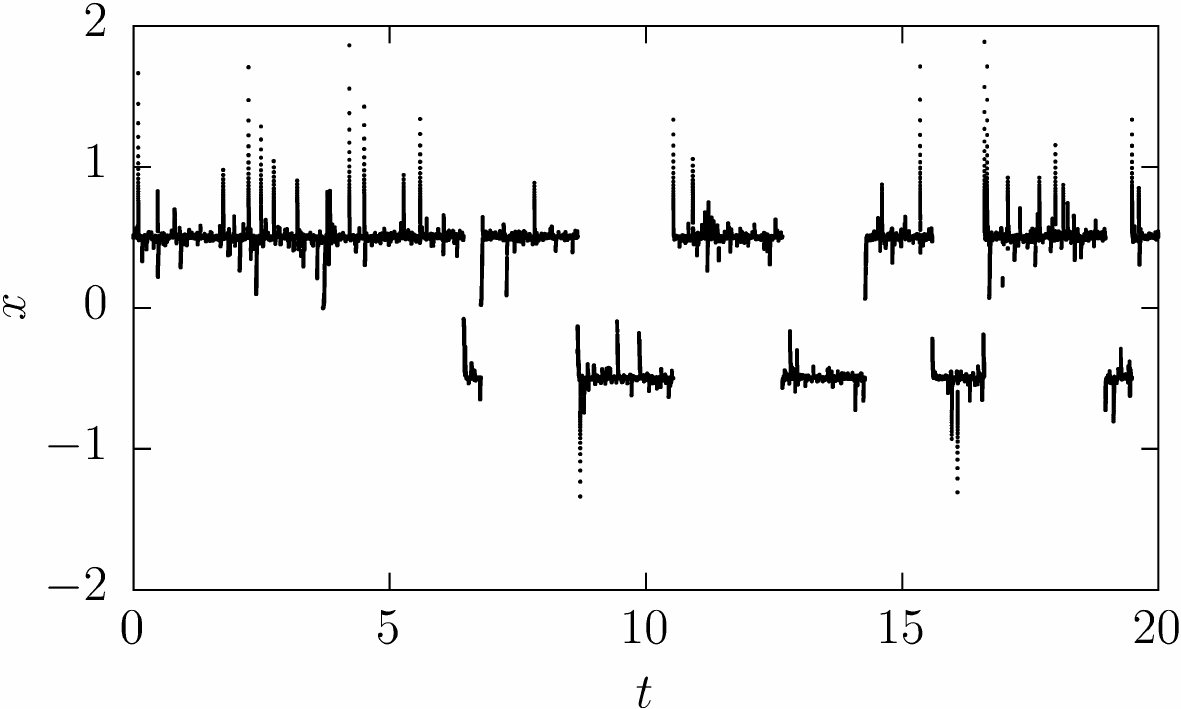}
\caption{
\bd{
Survival probability, i.e. complementary cumulative density of first passage times (top panel)  
and exemplary trajectories for $\alpha=2$ (middle panel) and $\alpha=1$ (bottom panel)
for a continuous potential~(\ref{eq:ciagly}) with $a=1$.
The trajectory for $\alpha=1$ has been plotted with symbols, in order to emphasise its discontinuity. 
}
}
\label{fig:fpt-trajectories}
\end{figure}

Top panel of Fig.~\ref{fig:zat_E85} presents results for $\Delta E_1=8.5$, $\Delta E_2=7.5$, $x_1=0.25$ and $x_2=0.75$. 
Orange dots depict the ratio ($\mathcal{P}=P_2/P_1$) of occupation probabilities whereas blue dots represent the ratio ($\mathcal{T}=T(x_1)/T(x_2)=k_{12}/k_{21}$) of transition rates.
The blue solid line shows the ``width ratio'' $(\delta_2/\delta_1)^\alpha$ predicted by Eq.~(\ref{eq:ditlevsen}), while the orange dashed line depicts the ``depth ratio'' $(\Delta E_1/\Delta E_2)^\alpha$ given by Eq.~(\ref{eq:bier}).
For small values of the stability index $\alpha$, blue dots and orange squares coincide, while for $\alpha \geqslant 1.3$ they start to differ.
The scaling predicted by Eq.~(\ref{eq:ditlevsen}) is confirmed by numerical simulations with $\alpha<1.3$.
In the top panel of Fig.~\ref{fig:zat_E85} there are two additional points.
There is a red triangle corresponding to the analytically calculated value of MFPTs for $\alpha=2$.
Moreover, using the stationary $p(x) \propto \exp(-V(x)/\sigma^2)$ representative for this case we can evaluate
$P_1$ and $P_2$ from Eqs.~(\ref{eq:p1}) and~(\ref{eq:p2}).

The green symbol (rhombus) in the top panel of Fig.~\ref{fig:zat_E85} indicates ratio $\mathcal{T}$ calculated from Eq.~(\ref{eq:mfpt}).
The ratios of reaction rates, calculated by use of exact formulas valid for $\alpha=2$, significantly differ from the prediction of ``width ratio'' given by Eq.~(\ref{eq:ditlevsen}), but they are close to the ``depth ratio'' scaling predicted by Eq.~(\ref{eq:bier}), see the red triangle and the green rhombus in the top panel of Fig.~\ref{fig:zat_E85}.


Relations given by Eqs.~(\ref{eq:full}) and the ratio $\mathcal{P}$  hold only when transitions between potential wells are performed in a single long jump \cite{ditlevsen1999anomalous,vezzani2018single}.
This condition is well satisfied in the limit of vanishing noise intensity for deep potential wells, when the particle is driven by the L\'evy noise with the small value of the stability index $\alpha$.
Contrary to small $\alpha$, for $\alpha$ large enough the central part of a noise distribution plays an increasingly important role, as for growing $\alpha$ more probability mass is located around $x=0$. 
If a slope of the potential barrier is not steep enough, multiple-step re-crossings of the barrier become more frequent \cite{ciesla2019multimodal}. 
Therefore, not only Eq.~(\ref{eq:population}) does not hold but also transitions from a shallower potential well become more probable.

The middle panel of  Fig.~\ref{fig:zat_E85} presents results for a deep potential well.
In contrast to the top panel of Fig.~\ref{fig:zat_E85}, there is an additional green dot-dashed line corresponding to the full formula given by Eq.~(\ref{eq:full}).
For a very deep potential well, the $\alpha$ dependence predicted by Eq.~(\ref{eq:full}) is the closest to the results of stochastic simulations.
It indicates existence of the regime where effects of the barrier width and barrier height contribute to the evaluated rate.
This regime corresponds to a finite discretization time step $\Delta t$ fulfilling the additional constraint
\begin{equation}
\delta \sim \Delta E \Delta t.    
\end{equation}
Otherwise, in the limit of $\Delta t \to 0$, the ``width ratio'' scaling predicted by Eq.~(\ref{eq:ditlevsen}) is visible.

The bottom panel of Fig.~\ref{fig:zat_E85} presents results for the potential (\ref{eq:ciagly})  with $a=1$ along with two lines corresponding to limiting scaling given by Eqs.~(\ref{eq:ditlevsen}) (blue solid line) and (\ref{eq:bier}) (orange dashed line).
For $a=1$ with $\alpha<2$, numerical results obtained by use of the ``in-well'' population method or the MFPT ($\mathcal{P}$ versus $\mathcal{T}$) are coherent.
Furthermore, for $\alpha<2$, calculated ratios are very close to the ``width ratio'' prediction of Eq.~(\ref{eq:ditlevsen}).
Note, that despite the approximation~(\ref{eq:ditlevsen}) is valid in the limit of vanishing noise \cite{ditlevsen1999anomalous,imkeller2006,imkeller2010hierarchy} it seems to work very well also for finite noise strength $\sigma$ \cite{chechkin2005,chechkin2007}.

\bd{
 The escape process is Markovian and characterized by finite MFPT, consequently first passage time distributions are exponential\cite{dybiec2007}. This characteristics can be observed by analysis of the survival probabilities depicted in 
 Fig.~\ref{fig:fpt-trajectories}.  Top panel of Fig.~\ref{fig:fpt-trajectories} presents sample survival probabilities (the probability that a particle remains within the initial potential well up to time $t$) 
under Cauchy ($\alpha=1$) and Gaussian ($\alpha=2$) drivings for escape events over the continuous potential given by Eq.~(\ref{eq:ciagly}) with $a=1$. 
Inspection of trajectories reveals difference in escape scenario induced by Gaussian and Cauchy noises.
Trajectories under action of the Gaussian noise are continuous and a particle surmounts the potential barrier in a series of subsequent jumps.
For the Cauchy driving, the trajectory is discontinuous and escape is typically performed in a single long jump. Moreover, for $\alpha=1$, a particle can make distant excursions to outer points.
Replacement of the continuous potential with the piece-wise liner, see top panel of Fig.~\ref{fig:linearpotential}, bounds the motion to the finite interval restricted by minima of the potential.
}





\subsection{Additive thermal and L\'evy noise \label{sec:mixture}}
The scaling of the ratio of escape rates, see Eqs.~(\ref{eq:full}) and (\ref{eq:ditlevsen}) is derived using the asymptotic properties of $\alpha$-stable densities.
Such a derivation disregard the central part of the random force distribution.
The central part of the jump length distribution control short jumps which are responsible for penetration of the potential barrier \cite{ditlevsen1999anomalous}.
Therefore, in the current subsection, we assume that the particle is driven by two stochastic forces \cite{ebeling2009microfields,ebeling2010convoluted,kusmierz2018,kusmierz2014}, so that the Langevin equation assumes the form
\begin{equation}
\frac{dx}{dt}=- V'(x) + \sigma \xi (t) + \eta (t).
\label{eq:langevin-twonoises}
\end{equation}
As in Eq.~(\ref{eq:langevin}), $\xi (t)$ stands for the  L\'evy noise whereas the additional, independent term $\eta (t)$ is assumed to be the Gaussian white noise, describing thermal fluctuations in the system.
In such a setup the Gaussian white noise can be considered as the internal noise, while the L\'evy noise is the external fluctuating force. 
\bd{Putting it differently, parameters of the Gaussian noise are defined (with help of fluctuation dissipation theorem) by internal parameters, while parameters of the L\'evy noise are externally controlled\cite{kusmierz2014,kanazawa2015minimal}.}
For the sake of clarity, from now on we assume that the intensity of the Gaussian fluctuations stays fixed,  i.e. it is set to unity.
The scale parameter $\sigma$ describes then the strength of the external L\'evy noise with respect to the intensity of thermal fluctuations.
As the reference case for the study of a combined action of two independent noise sources, we use the continuous potential of Sec.~\ref{sec:plp}, see Eq.~(\ref{eq:ciagly}) and bottom panel of Fig.~\ref{fig:linearpotential}.
Therefore, we use the same potential as in the bottom panel of Fig.~\ref{fig:zat_E85}, i.e. the potential given by Eq.~(\ref{eq:ciagly}) with $a=1$ or $a=10$.
Please note, that the model studied in bottom panel of Fig.~\ref{fig:zat_E85} corresponds to Eq.~(\ref{eq:langevin-twonoises}) with $\eta(t)\equiv 0$ and $a=1$.

\begin{figure}[h!]
\centering
\includegraphics[width=0.45\textwidth]{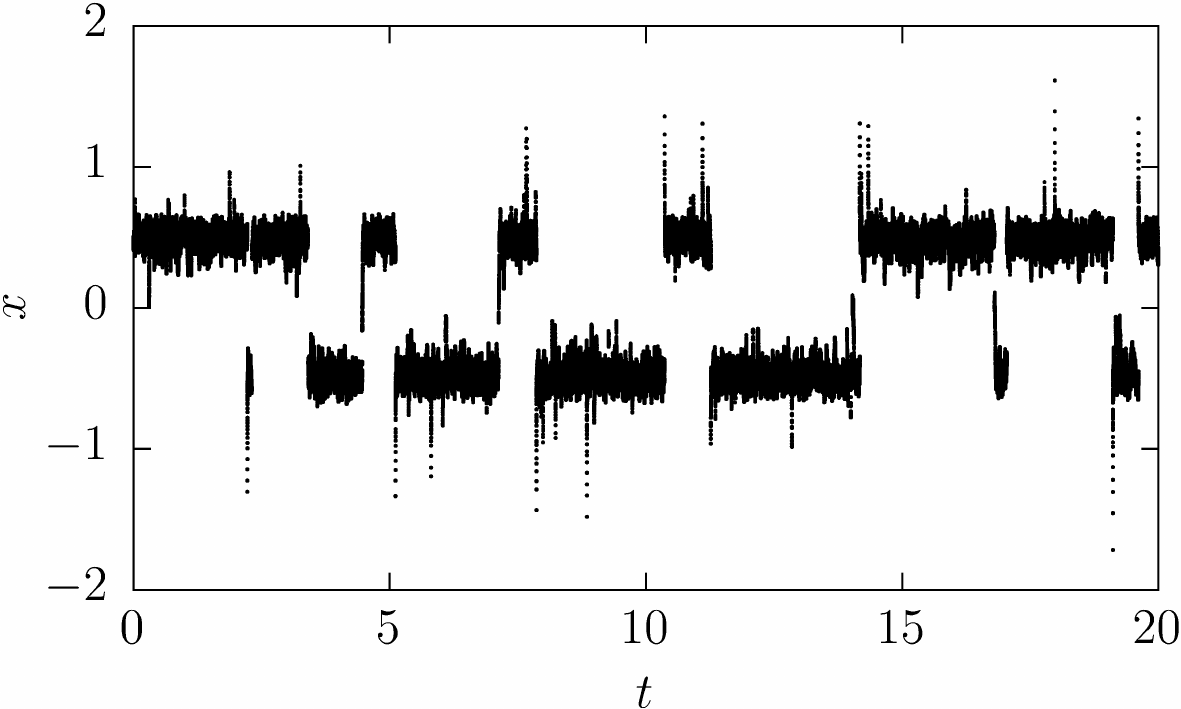}\\
\caption{
\bd{
Sample trajectory for a particle moving in the continuous double-well potential (\ref{eq:ciagly}) with $a=1$ driven by simultaneous action of Cauchy ($\alpha=1$) and Gaussian ($\alpha=2$) noises. 
}
}
\label{fig:trajectories2}
\end{figure}

\bd{
First, we verify how the combined action of two noises changes properties of trajectories.
Fig.~\ref{fig:trajectories2} presents a sample trajectory for a particle moving in a potential (\ref{eq:ciagly}) with $a=1$ driven by simultaneous action of Cauchy and Gaussian noise. 
In contrast to pure Cauchy driving, see bottom panel of Fig.~\ref{fig:fpt-trajectories}, trajectory explores more vicinity of potential's minima.
Moreover, due to Gaussian component of the stochastic driving a particle is more likely to visit neighborhood of the potential barrier.
Nevertheless, majority of escape event is still performed in a single long jump, but now the last visited point before escape from a potential well is typically closer to the boundary than for pure Cauchy driving.
}

\begin{figure}[h!]
\centering
\includegraphics[width=0.9\columnwidth]{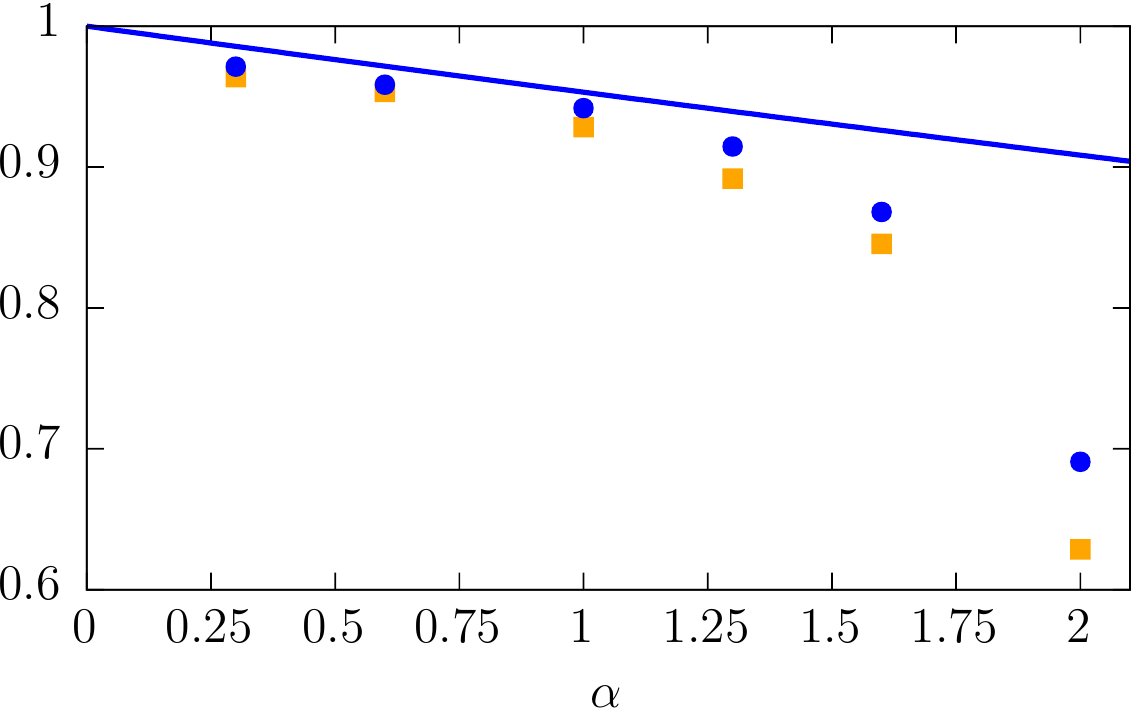} \\
\includegraphics[width=0.9\columnwidth]{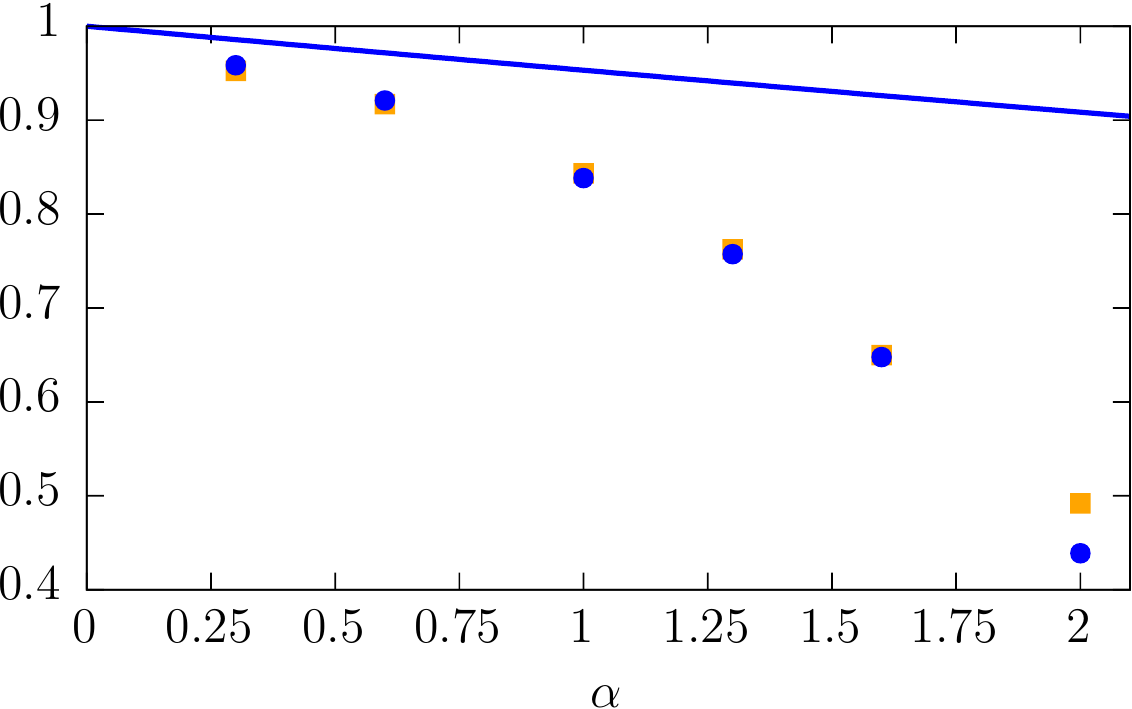} \\
\includegraphics[width=0.9\columnwidth]{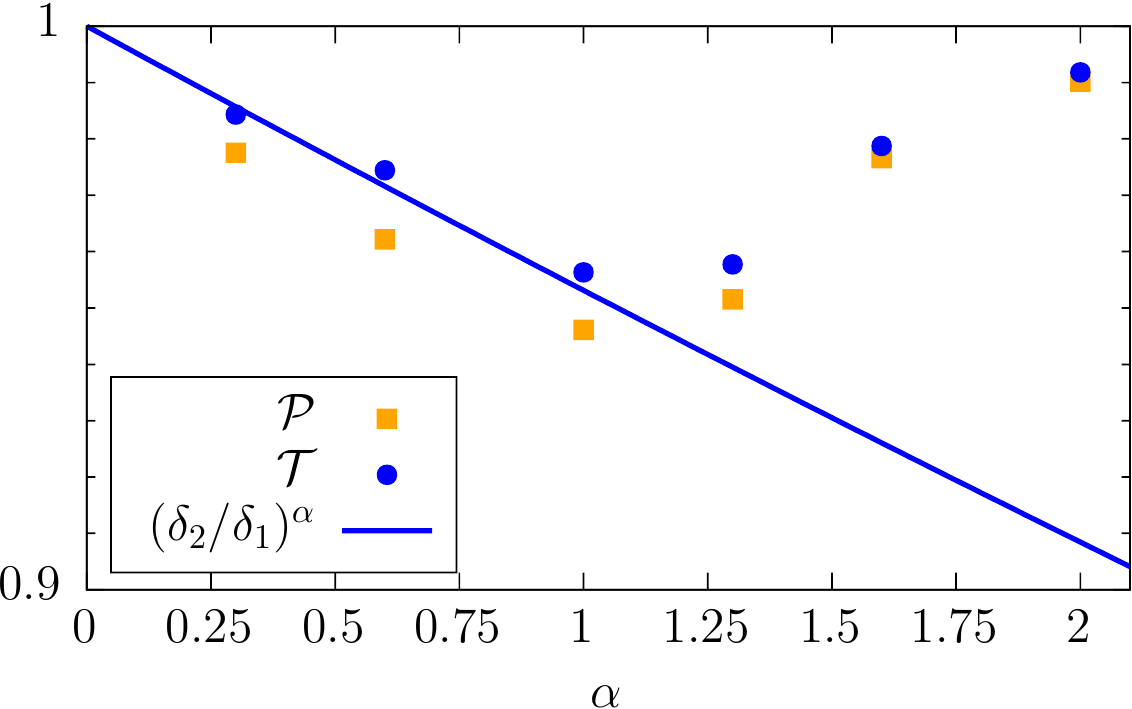} \\
\caption{
Symbols represent the ratios $\mathcal{P}$ of occupation probabilities (\textcolor{orange}{$\blacksquare$}) and $\mathcal{T}$ of transition rates  (\textcolor{blue}{$\bullet$})
for the continuous double-well potential (\ref{eq:ciagly}) with $a=1$. 
Solid lines show  the theoretical ``width ratio'' scaling (blue solid, see Eq.~(\ref{eq:ditlevsen})). 
%
%
Subsequent panels correspond to various values of the $\sigma$  parameter scaling the strength of L\'evy noise: 
$\sigma=1$ (top panel),
$\sigma=0.1$ (middle panel) and 
$\sigma=10$ (bottom panel).
The legend is included in the bottom panel.}
\label{fig:tl_s1}
\end{figure}

Comparison of bottom panel of Fig.~\ref{fig:zat_E85} and top panel of Fig.~\ref{fig:tl_s1} reveals that incorporation of the additional Gaussian noise source significantly weakens the agreement between the prediction of ``width ratio'' given by Eq.~(\ref{eq:ditlevsen}) and results of computer simulations.
The presence of the Gaussian noise changes the escape scenario by increasing chances of an escape in a sequence of jumps \cite{bier1998adiabaticity}.
Consequently, due to the increased width of the central part of the jump length distribution, the role played by the tails of L\'evy distribution is depleted, what in turn results in stronger deviations from the ``width ratio'' predicted by Eq.~(\ref{eq:ditlevsen}).
These deviations are amplified for all values of the stability index $\alpha$, except the special case of $\alpha=2$. For $\alpha=2$, the L\'evy noise is equivalent to the Gaussian white noise.
Therefore, the presence of two Gaussian noise sources facilitate escape kinetics as they can be combined in the single Gaussian white noise with the increased width.
\bd{The increased width of the resultant Gaussian white noise, with help of fluctuation dissipation theorem, can be attributed to the increase in the system temperature.
Please, note that the situation is more subtle for the underdamped models, which are not studied here.}
For the potential given by Eq.~(\ref{eq:ciagly}) with $a=1$, the scaling predicted by Eq.~(\ref{eq:ditlevsen}) is very similar to the ratio given by Eq.~(\ref{eq:gauss}).

In the middle panel of Fig.~\ref{fig:tl_s1}, the scale parameter $\sigma$ is reduced to $\sigma=0.1$.
For lower $\sigma$ the central part of the jump length distribution, amplified due to presence of the Gaussian white noise, is even more prominent.
In the middle panel of Fig.~\ref{fig:tl_s1}, deviations between the weak noise theory, see Eq.~(\ref{eq:ditlevsen}) and the actual scaling are amplified.
In order to assure that the increased disagreement is due to presence of the Gaussian component we have performed additional simulations with $\sigma=0.1$ and $\eta \equiv 0$.
For $\sigma=0.1$ and $\eta \equiv 0$, we obtained results which are quantitatively indistinguishable from those one included in  the bottom panel of Fig.~\ref{fig:zat_E85}.
This effect indicates that the increased disagreement in the middle panel of Fig.~\ref{fig:tl_s1} is produced by the action of the Gaussian\bd{, thermal} white noise.
Moreover, it demonstrates that the approximation given by Eq.~(\ref{eq:ditlevsen}), which is derived in the $\sigma\to 0$ limit, works pretty well for finite $\sigma$, see bottom panel of Fig.~\ref{fig:zat_E85}. 

The bottom panel of Fig.~\ref{fig:tl_s1} examines the model for $\sigma=10$. 
The agreement between results of computer simulations and Eq.~(\ref{eq:ditlevsen}) seems to be restored. 
Unfortunately, this agreement is a coincidence due to the potential shape and the combined action of two very different effects.
First of all, tails of the jump length distribution leads to the scaling predicted by Eq.~(\ref{eq:ditlevsen}).
Nevertheless, due to a large value of the scale parameter $\sigma$, also the central part of the jump length distribution becomes non-negligible.
The influence of the central part of the jump length distribution on the escape kinetics can be quantified by the MFPT for a system driven by a Gaussian noise with some effective \cite{ditlevsen1999anomalous} $\sigma_{\mathrm{eff}}$.
Due to the shape of the potential, more precisely almost symmetric location of potential's minima, both scalings (\ref{eq:ditlevsen})  and~(\ref{eq:gauss}) give similar approximations for the ratio of reaction rates.

\begin{figure}[h!]
\centering
\includegraphics[width=0.9\columnwidth]{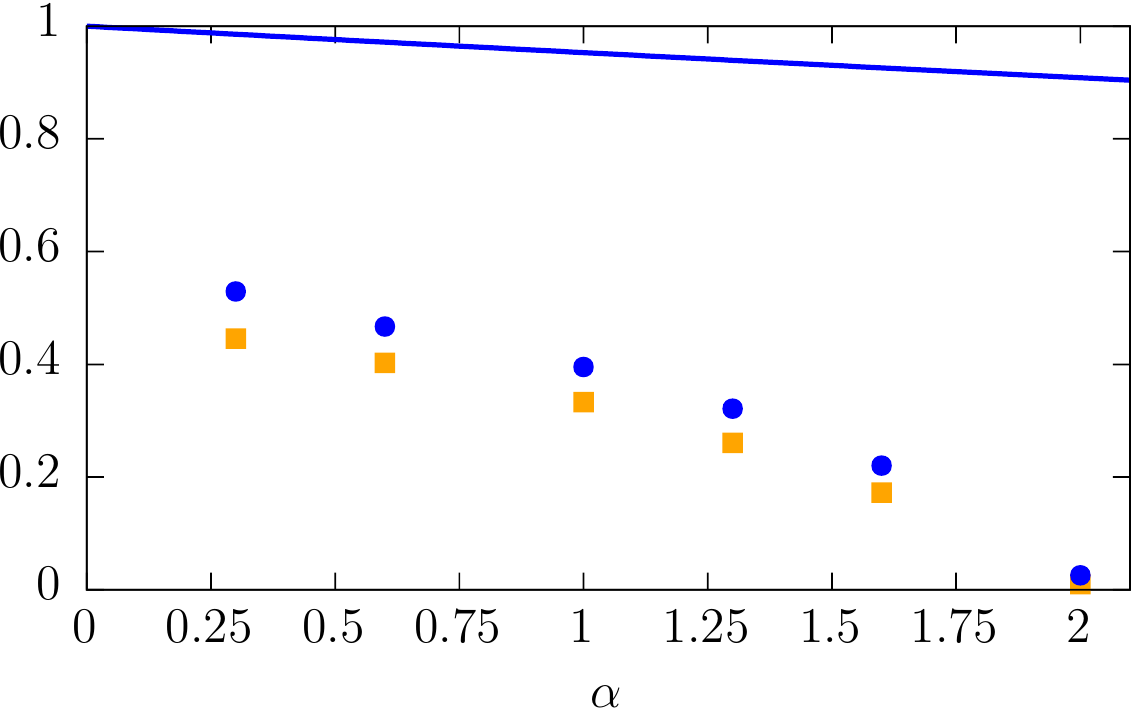} \\
\includegraphics[width=0.9\columnwidth]{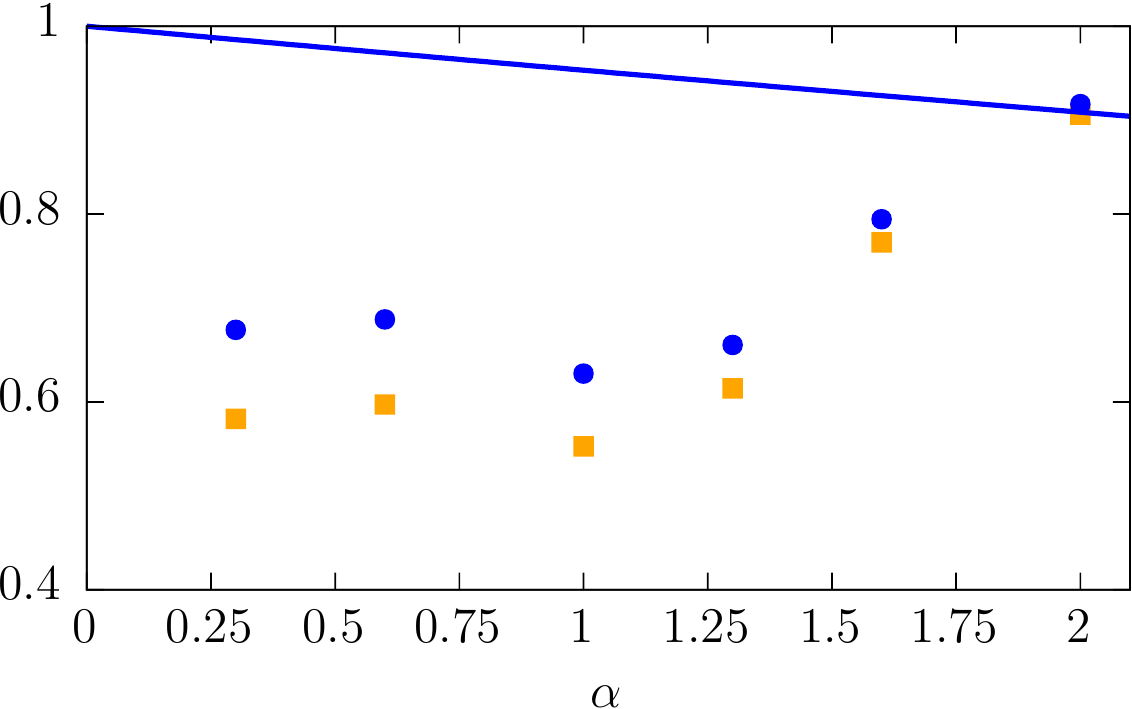}\\
\includegraphics[width=0.9\columnwidth]{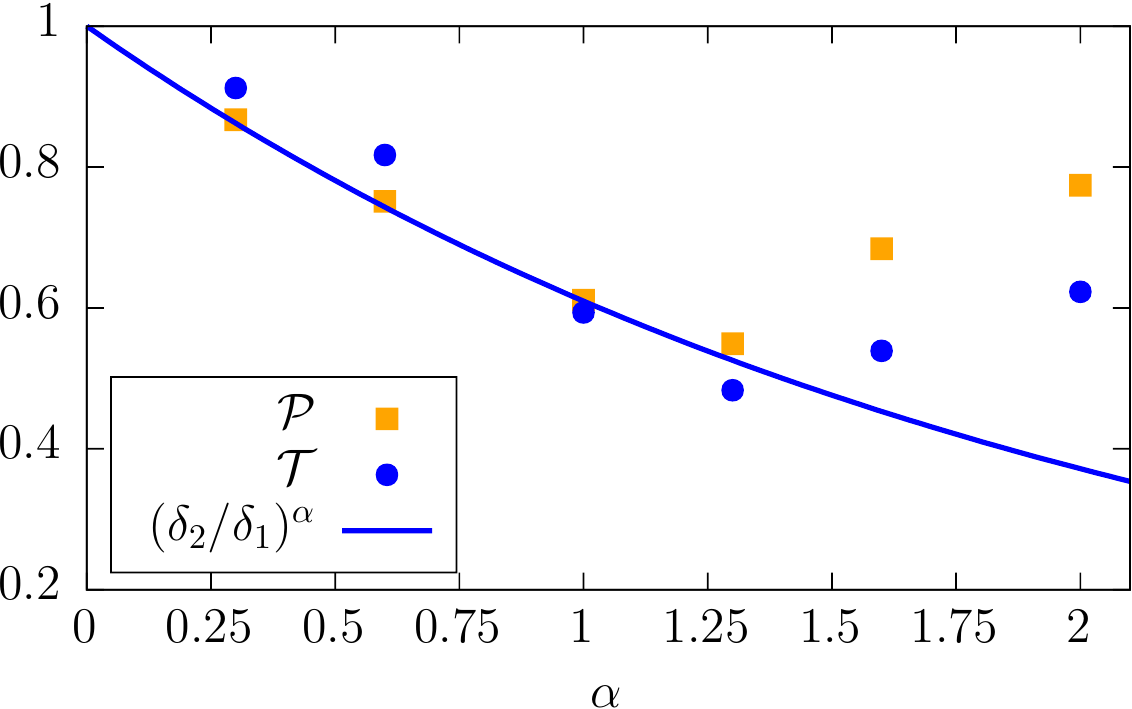}\\
\caption{The same as in Fig.~\ref{fig:tl_s1} for $a=10$ with
$\sigma=1$ (top panel),
$\sigma=10$ (middle panel) and
 $\sigma=10$ with $\eta(t)\equiv 0$  (bottom panel).
The legend is included in the bottom panel.
}
\label{fig:tl_s1a10}
\end{figure}

To eliminate this accidental agreement, the potential (\ref{eq:ciagly}) with $a=10$ was used. 
Now minima of the potential have depth of $\Delta E_1 \approx 13.2$ and $\Delta E_2 \approx 3.2$ and their locations are not as symmetric as for $a=1$, see bottom panel of Fig.~\ref{fig:linearpotential}.
As it is clearly visible in Fig.~\ref{fig:tl_s1a10}, the results of computer simulations with $a=10$ and $\sigma=1$ differ from Eq.~(\ref{eq:ditlevsen}).
The pronounced disagreement  is produced by the Gaussian white noise component, which increases likelihood of continuous (instead of single jump) transition over the potential barrier.
Furthermore, for $a=10$, the right potential well is shallow, what further increases deviations from the scaling given by Eq.~(\ref{eq:ditlevsen}).
In the middle panel of Fig.~\ref{fig:tl_s1a10} the scale parameter $\sigma$ is increased to $\sigma=10$.
Paradoxically, the disagreement between the asymptotic scaling and results of computer simulations, due to a presence of the Gaussian white noise, is further amplified by the L\'evy noise.
More precisely, for $\sigma=10$, the assumption of a weak noise, which is crucial for asymptotics predicted by  Eq.~(\ref{eq:ditlevsen}), does not hold, even without the Gaussian white noise.

The agreement between results of computer simulations and the asymptotic scaling~(\ref{eq:ditlevsen}) can be reintroduced by disregarding the Gaussian white noise source, i.e. by setting $\eta(t)\equiv 0$ as in the bottom panel of Fig.~\ref{fig:tl_s1a10}. 
For instance, for $a=10$ with $\sigma=1$ the agreement is significant (results not shown).
At the same time for the increased $\sigma=10$ the accordance is observed for $\alpha<1$, see bottom panel of Fig.~\ref{fig:tl_s1a10}.
For $\alpha>1$ with $\sigma=10$ the central part of the L\'evy distribution is too wide to make the single jump escape scenario dominating what in turn introduces violations of Eq.~(\ref{eq:ditlevsen}).

From the examination of the escape kinetics driven by the combined action of two independent L\'evy and Gaussian noise sources we can deduct following scenarios resulting in the violation of ``width ratio'' given by Eq.~(\ref{eq:ditlevsen}): (i) addition of the Gaussian white noise source, (ii) increasing of the scale parameter in the L\'evy noise and (iii) decreasing depth of potential wells.
The scenarios (i) and (ii) are related, because both of them increase the width of the central part of the jump length distribution which is responsible for the penetration of the potential barrier, see Ref.~\onlinecite{ditlevsen1999anomalous}.
Consequently, elimination of the Gaussian noise, under the condition that $\sigma$ is small enough, reintroduces the scaling given by Eq.~(\ref{eq:ditlevsen}), see Fig.~\ref{fig:zat_E85} and bottom panel of Fig.~\ref{fig:tl_s1a10}.
Nevertheless, due to finite $\sigma$, when $\alpha \to 2$ even in Fig.~\ref{fig:zat_E85} and bottom panel of Fig.~\ref{fig:tl_s1a10} violations of Eq.~(\ref{eq:ditlevsen}) are visible.
These violations can be decreased by the reducing the scale parameter $\sigma$.
Finally, the scenario (iii) breaks the two state approximation as ``in-well'' densities become wide.

\begin{figure}[h!]
\centering
\includegraphics[width=0.98\columnwidth]{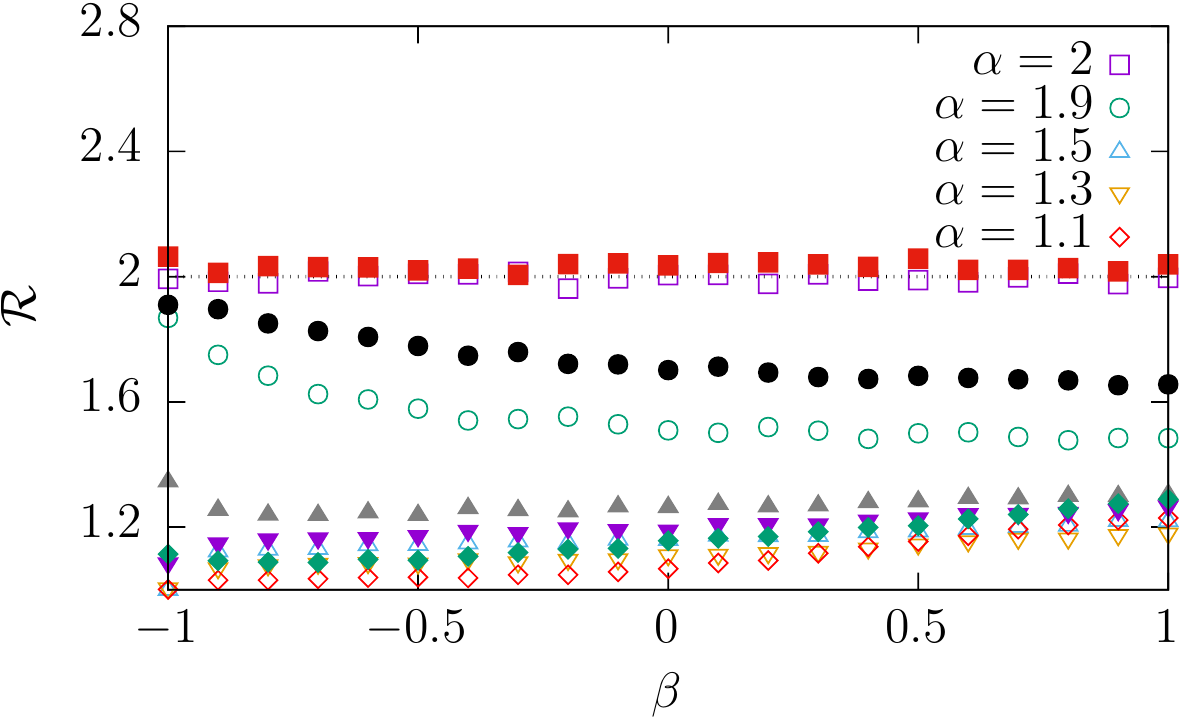}
\caption{
Ratio $\mathcal{R}$ of mean first passage times for well-bottom-to-well-bottom $T_{w-w}$ and well-bottom-to-barrier-top $T_{w-b}$ for the L\'evy noise (empty points) and mixture of Gaussian and L\'evy noises (full symbols) for the symmetric double-well potential given by Eq.~(\ref{eq:ciagly}) with $a=0$.
For more details see~Ref.~\onlinecite{dybiec2007}.
}
\label{fig:ratio}
\end{figure}

In the Ref.~\onlinecite{dybiec2007}, we have studied the model of escape kinetics induced by general $\alpha$-stable noises in a symmetric double-well potential given by Eq.~(\ref{eq:ciagly}) with $a=0$.
In particular, for a particle starting in one of the potential wells we have calculated the ratio $\mathcal{R}=T_{w-w}/T_{w-b}$ of mean first passage times for well-bottom-to-well-bottom $T_{w-w}$ and well-bottom-to-barrier-top $T_{w-b}$ escape scenarios.
For the Gaussian white noise such a ratio is equal to two, i.e. $\mathcal{R}=2$, see \onlinecite{hanggi1990}.
Action of the L\'evy noise breaks this property of escape kinetics in double-well potentials --- the ratio of MFPTs becomes smaller than two.
Addition of the Gaussian white noise  (with the scale parameter set to unity)  to the model considered in Ref.~\onlinecite{dybiec2007} increases the value of the ratio of mean first passage times approximately by 10\%.
The ratio has increased because the additional Gaussian white noise increased the role played by the central part of the jump length distribution.
Nevertheless, the ratio $\mathcal{R}$ is still smaller than two, see Fig.~\ref{fig:ratio}.
For more details see~Ref.~\onlinecite{dybiec2007}.

\subsection{Escape from finite intervals \label{sec:finite}}

\begin{figure}[!h]%
\centering
\begin{tabular}{c}
\includegraphics[width=0.85\columnwidth]{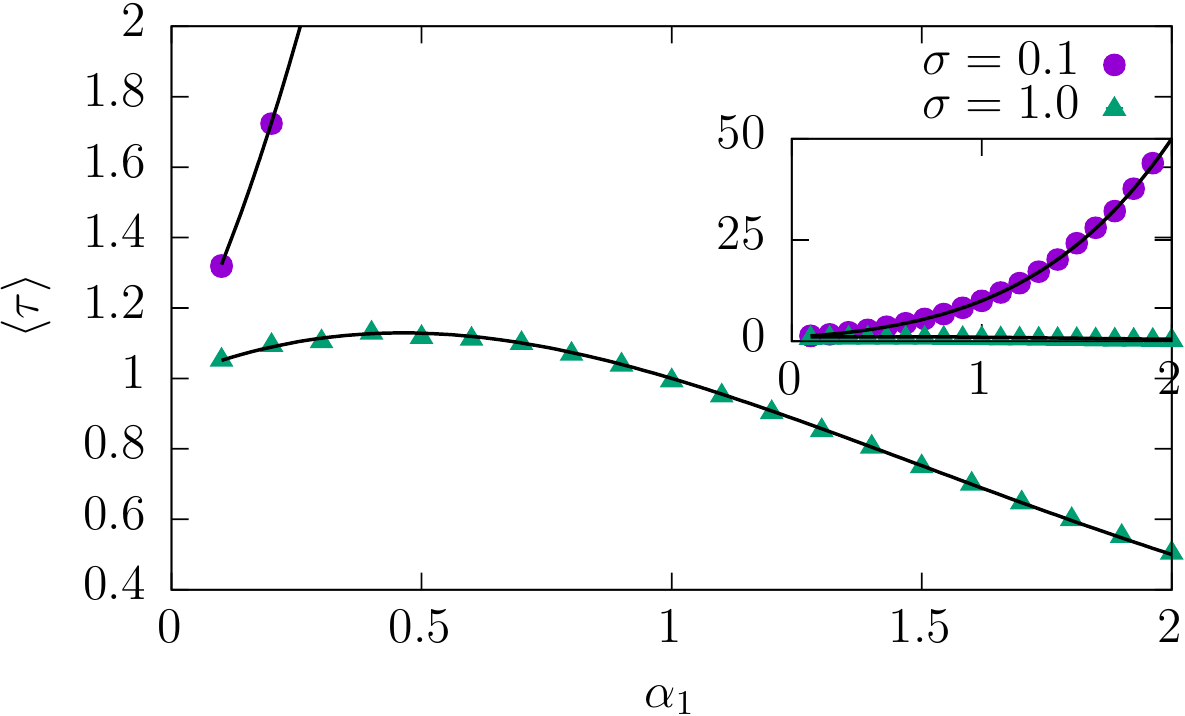} \\
\includegraphics[width=0.85\columnwidth]{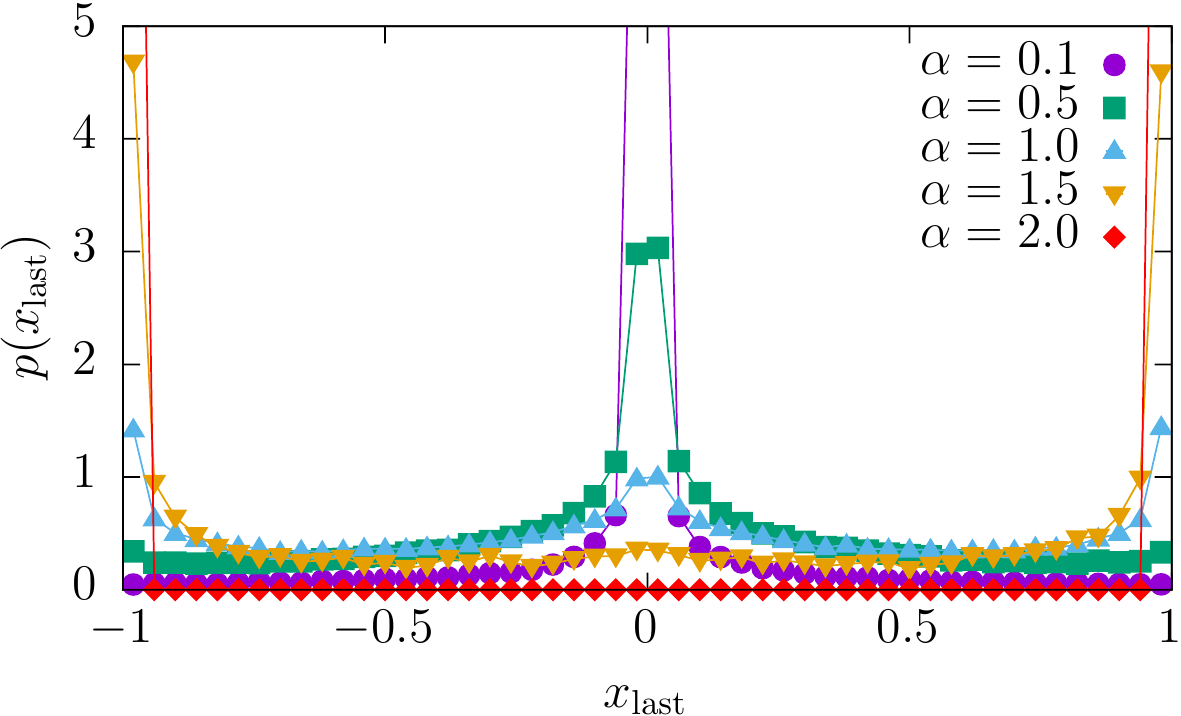} 
\end{tabular}
\caption{
Mean first passage time $\langle \tau \rangle$ for the escape from the finite interval $[-1,1]$ (top panel) and the last hitting point density $p(x_{\mathrm{last}})$ for $\sigma=0.1$ (bottom panel).
Solid lines in the top panel correspond to the exact, theoretical formula, see Eq.~(\ref{eq:mfpt}). Other parameters: initial condition $x(0)=0$,  time step of integration $\Delta t=10^{-4}$ and number of repetitions $N=10^5$. }
\label{fig:mfpt-sn}
\end{figure}

From the examination of the escape kinetics induced by a mixture of noises it can be deducted that addition of thermal noise changes the escape kinetics.
Presence of the additional thermal noise  changes the escape protocol from a single long jump scenario to a sequence of shorter jumps controlled by the central part of the jump length distribution.
In order to elucidate this issue in more details, we switch to the archetypal model of escape from the finite interval $[-L,L]$.
Initially a particle is located in the middle of the interval, i.e. $x(0)=0$, and the motion is continued until $|x|<L$.
The exact formula for the MFPT reads
\begin{equation}
    \langle \tau \rangle = \frac{1}{\Gamma(1+\alpha)} \frac{L^\alpha}{\sigma^\alpha},
    \label{eq:mfpt-interval}
\end{equation}
see Refs.~\onlinecite{blumenthal1961, getoor1961, kac1950distribution, widom1961stable, kesten1961random}.
Furthermore, the intuitive argumentation supporting Eq.~(\ref{eq:mfpt-interval}) is included in the Appendix~\ref{app:mfpt}.
In addition to the MFPT, we studied the last hitting point densities $p(x_{\mathrm{last}})$, where $x_{\mathrm{last}}$ is the last point visited before leaving the $[-1,1]$ interval.
Top panel of Fig.~\ref{fig:mfpt-sn} presents MFPT as a function of the stability index $\alpha$ for $L=1$ with $\sigma=1$ and $\sigma=0.1$.
Results of computer simulations nicely follow theoretical curve given by Eq.~(\ref{eq:mfpt-interval}) with $L=1$.
Results for $\sigma=1$ are presented in the main plot, while for $\sigma=0.1$ in the inset, as values of MFPT for $\sigma=0.1$ are significantly larger than for $\sigma=1$. 
For $\sigma=0.1$ escape kinetics slows down with the increase of the stability index $\alpha$ because $(L/\sigma)^\alpha=(1/0.1)^\alpha=10^\alpha$ is a growing function of the stability index $\alpha$, see Eq.~(\ref{eq:mfpt-interval}).

The bottom panel of Fig.~\ref{fig:mfpt-sn} shows the last hitting point density for $\sigma=0.1$.
The $p(x_{\mathrm{last}})$ distribution for $\sigma=1.0$ is practically the same as for $\sigma=0.1$, therefore we show the distribution for $\sigma=0.1$ only.
For processes with continuous trajectories $x_{\mathrm{last}}=\pm L$ because the escape is performed by approaching of one of the absorbing boundaries. The very different situation is observed for L\'evy flights, which have discontinuous trajectories.
The most probable $x_{\mathrm{last}}$ is the origin, as $x(0)=0$, but with the increasing $\alpha$, maxima at the borders emerge.
The escape from the vicinity of the initial position can be dominating, but the escape itself is it not immediate.
For example, for $\alpha=0.5$, on average the escape occurred after approx $10^4$ jumps since $\Delta t=10^{-4}$.
Initial short jumps (controlled by the central part of the jump length distribution) resulted in the spreading of the last visited point around the initial condition.
Bottom panel of Fig.~\ref{fig:mfpt-sn} confirms that, for small values of the stability index $\alpha$ ($\alpha<1$), the escape from the vicinity of the initial position is the most probable. 
The different situation is observed  for $\alpha>1$ when the random walker is very likely to approach absorbing boundaries.

In the next step, using the model of L\'evy noise induced escape, we study the differences between escape protocols for single noise and mixture of noises induced escape.
We use mixture of two L\'evy noises characterized by stability indices $\alpha_1$ and $\alpha_2$ with $\sigma_1=\sigma_2=1$ or $\sigma_1=1,\;\sigma_2=0.1$.
Mixture of two independent L\'evy noises can be replaced by a single L\'evy noise if only they are characterized by the same stability index $\alpha$.
For $\alpha_1=\alpha_2$,  the sum of two independent identically distributed $\alpha$-stable random variables is distributed according to the $\alpha$-stable density with the same $\alpha$ and the scale parameter
\begin{equation}
    \sigma= \left [ \sigma_1^\alpha+\sigma_2^\alpha \right]^{\nicefrac{1}{\alpha}},
    \label{eq:sum-sigma}
\end{equation}
Therefore, using Eq.~(\ref{eq:mfpt-interval}) with $\sigma$ given by Eq.~(\ref{eq:sum-sigma}) it is possible to calculate the exact value of the MFPT.

\begin{figure}[h]%
\centering
\begin{tabular}{cc}
\includegraphics[width=0.45\columnwidth]{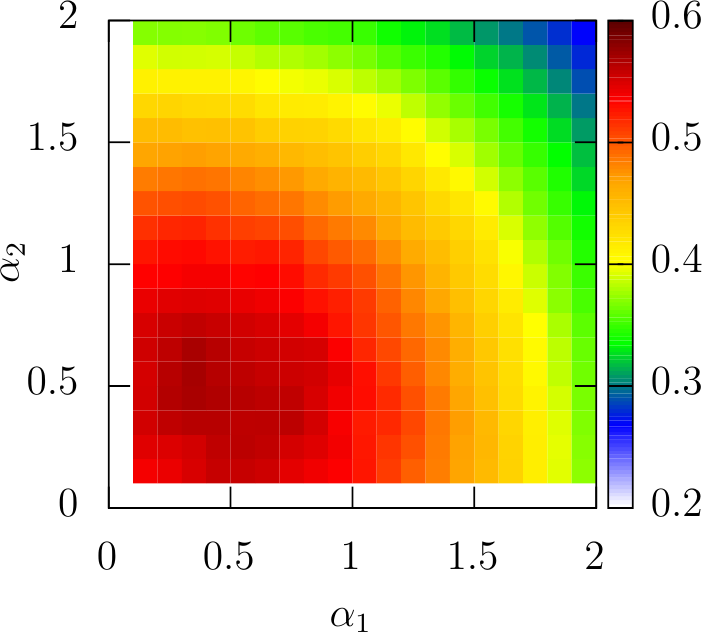} & \includegraphics[width=0.45\columnwidth]{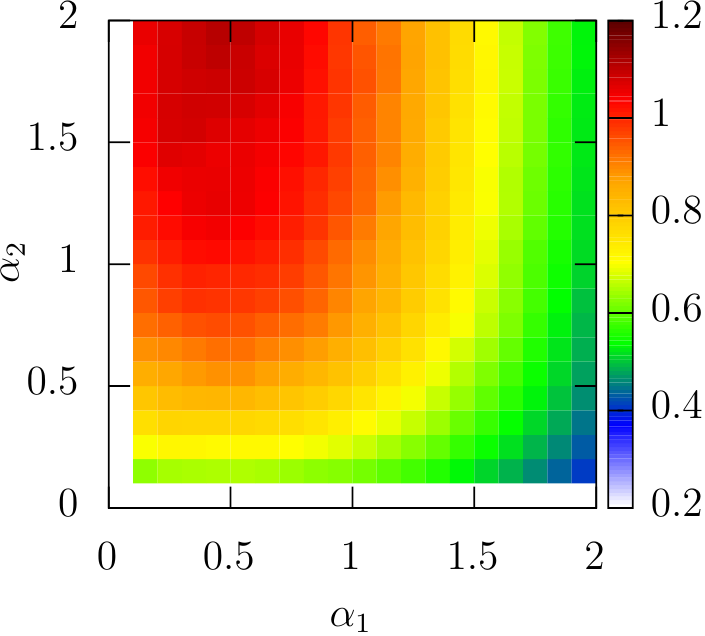}\\
\includegraphics[width=0.44\columnwidth]{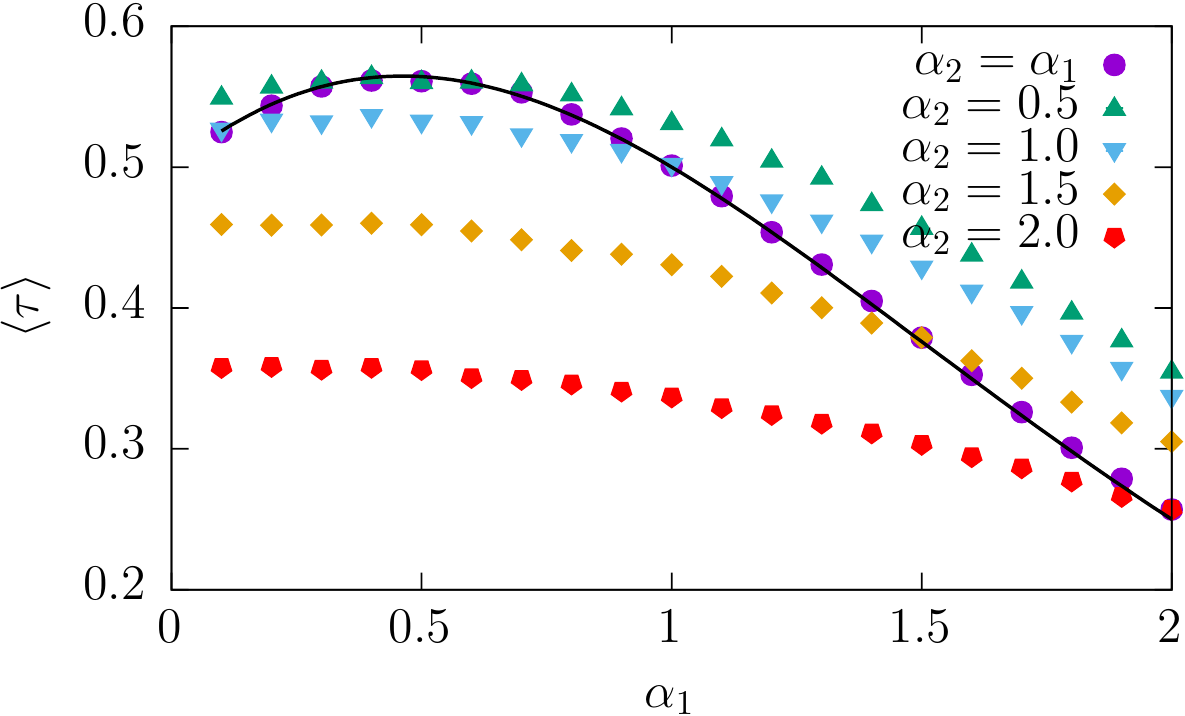} & \includegraphics[width=0.44\columnwidth]{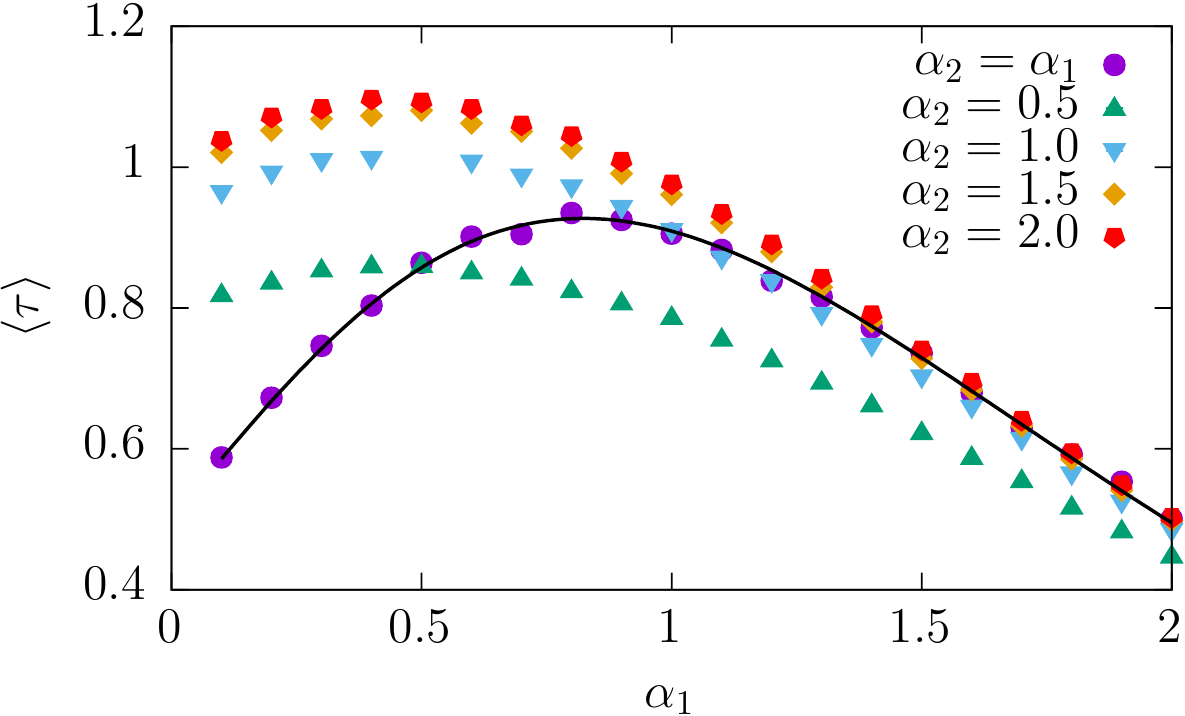}\\
\includegraphics[width=0.44\columnwidth]{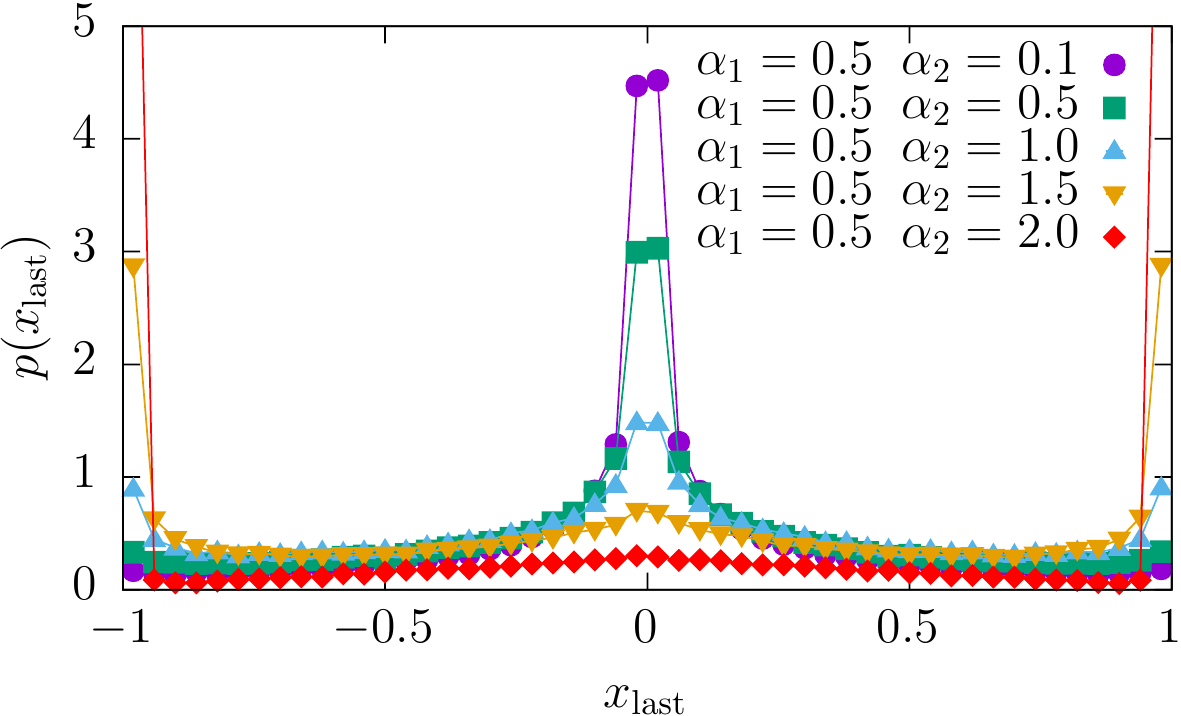} & \includegraphics[width=0.44\columnwidth]{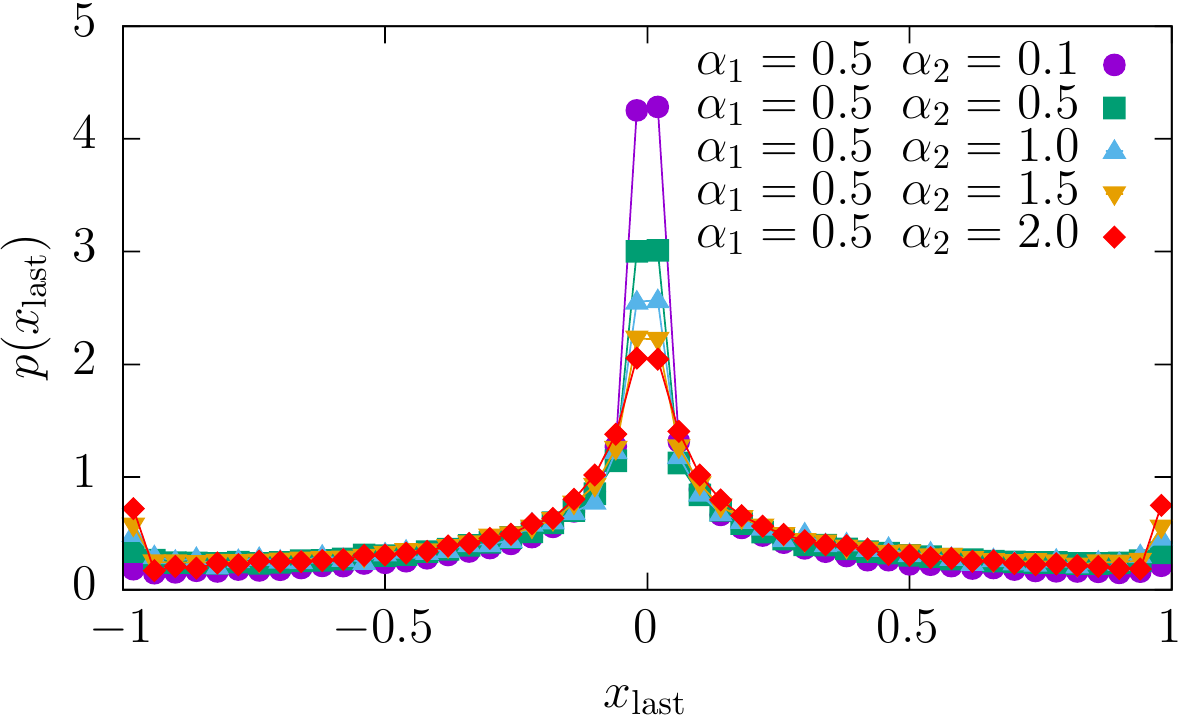} \\
\includegraphics[width=0.44\columnwidth]{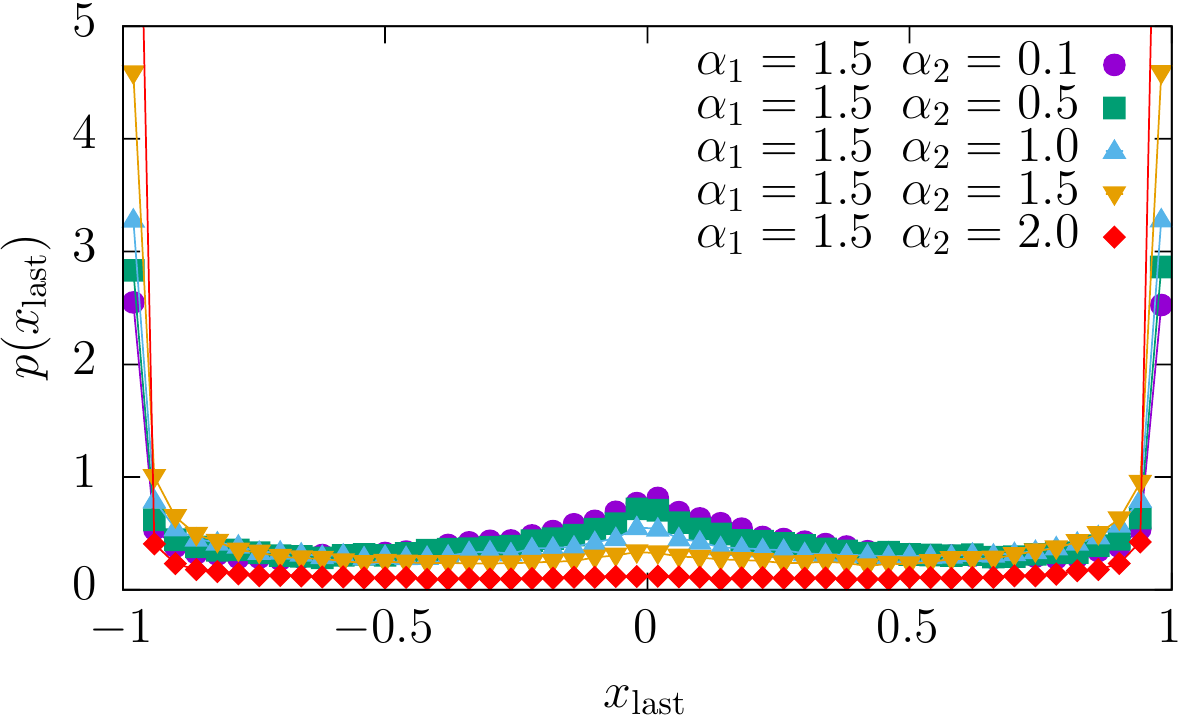} & \includegraphics[width=0.44\columnwidth]{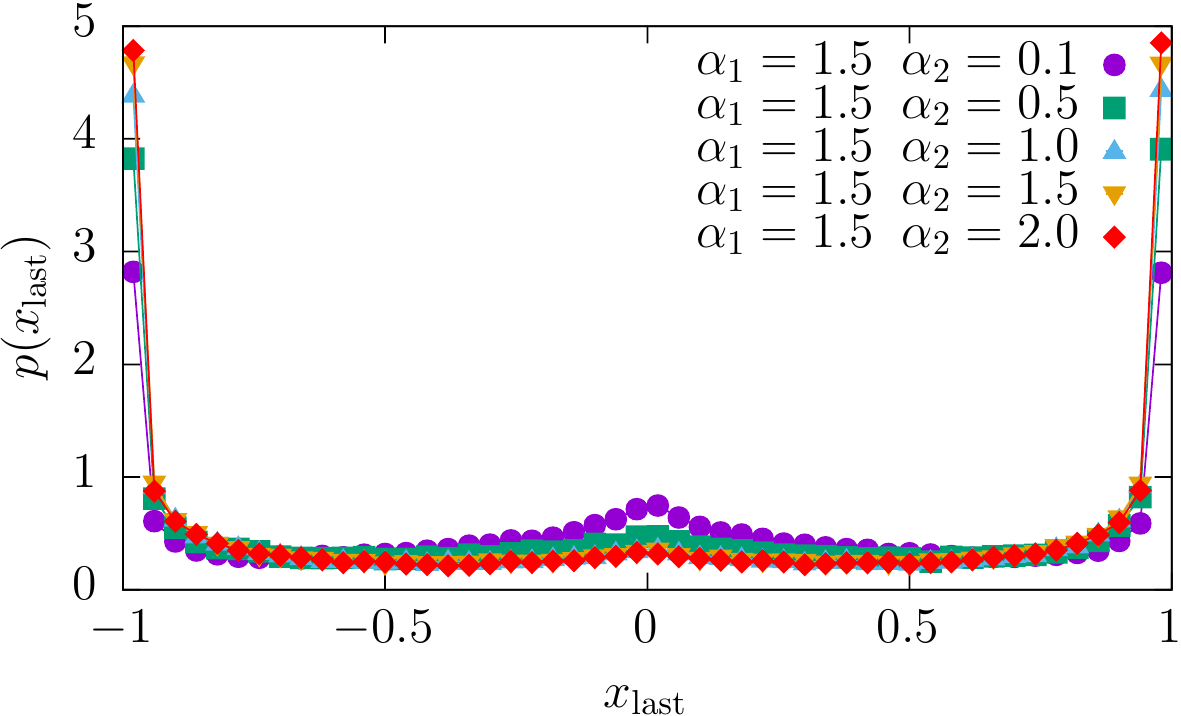} \\
\end{tabular}
\caption{
Mean first passage time $\langle \tau \rangle$ for the escape from the finite interval $[-1,1]$ (top panel), cross-section of the MFPT$(\alpha_1,\alpha_2)$ surface (second from the top panel),  and the last hitting point densities $p(x_{\mathrm{last}})$ (bottom panels).
Solid lines in the second from the top panel correspond to the exact, theoretical formula, see Eq.~(\ref{eq:mfpt-interval}). Other parameters: initial condition $x(0)=0$,  time step of integration $\Delta t=10^{-4}$ and number of repetitions $N=10^5$. 
Columns correspond to various values of the scale parameters: $\sigma_1=\sigma_2=1$ (left column) and $\sigma_1=1,\;\sigma_2=0.1$ (right column).
}
\label{fig:mfpt-tn}
\end{figure}

Figure~\ref{fig:mfpt-tn} presents results for escape driven by two noises. Subsequent columns correspond to different values of scale parameters: $\sigma_1=\sigma_2=1$ (left column) and $\sigma_1=1,\;\sigma_2=0.1$ (right column).
Top panel presents MFPT as function of stability indices $\alpha_1$ and $\alpha_2$.
Second from the top panel show sample cross-section of the MFPT surface.
For $\alpha_1=\alpha_2$ results of computer simulations (points) nicely follow exact results (solid lines), see Eqs.~(\ref{eq:mfpt-interval}) and~(\ref{eq:sum-sigma}).
Finally, bottom panels depict last hitting point densities $p(x_{\mathrm{last}})$ with $\alpha_1=0.5$ and $\alpha_1=1.5$.
For $\sigma_1=\sigma_2$, the MFPT surface is symmetric with respect to the interchange of $\alpha_1$ and $\alpha_2$, otherwise it is not symmetric along the diagonal.
For $\sigma_2=0.1$, the escape is slower because the width of the jump length distribution is reduced in comparison to $\sigma_2=1$, compare left and right panels of Fig.~\ref{fig:mfpt-tn}.

Examination of the last hitting point density  shows that addition of the second noise can modify the escape scenario.
For example, for $\alpha=0.5$, the most probable is escape from the vicinity of the initial position, see bottom panel of Fig.~\ref{fig:mfpt-sn}.
The bottom panel of Fig.~\ref{fig:mfpt-sn} should be contrasted with the second from the bottom panel of Fig.~\ref{fig:mfpt-tn} which present last hitting point densities for $\alpha_1=0.5$.
First of all, addition of the thermal noise (L\'evy noise with $\alpha=2$), produce peaks at boundaries both for $\sigma_2=1$ and $\sigma_2=0.1$, although for $\sigma_2=0.1$  their height is lower.
For $\sigma_2=1$ already addition of L\'evy noise with $\alpha>1$ produces modes at boundaries, while for $\sigma_2=0.1$ the first noise significantly  weakens the action of the second one. 

Bottom panels of Fig.~\ref{fig:mfpt-tn} presents last hitting point densities $p(x_{\mathrm{last}})$ for a free particle in finite interval, while the model studied in Secs.~\ref{sec:plp} and~\ref{sec:mixture} correspond to the motion in double-well potentials.
Nevertheless, already examination of the free motion is very instructive. 
It clearly shows that trajectories become more continuous-like with addition of the second noise with a larger value of the stability index $\alpha$.
Contrary to the free motion, in the case of external force, emergence of peaks at boundaries will be weakened because there is an external, deterministic, force pushing particles back to the potential minimum.
Moreover, due to the outer part of the potential, on the outer side of minima  particles experience the restoring force pushing them back to  minima of the potential.
This in turn increases the fraction of escape events from the potential minima, i.e. it amplifies $p(x_{\mathrm{last}})$ at $x_{\mathrm{last}} \approx x_1$ and $x_{\mathrm{last}} \approx x_2$.

\onecolumngrid

\clearpage

\twocolumngrid

%
%
%
\clearpage
\section{Summary and Conclusions\label{sec:summary}}

The noise induced escape over a potential barrier is an archetypal process modeling many phenomena.
In particular, it is a key element of the Kramers theory of chemical kinetics.
According to the Kramers theory, the reaction rate depends primarily on the relative height $\Delta E$ of the potential barrier separating states, $k \propto \exp(-\beta \Delta E)$, and decreases with the increasing barrier height. 
Such a dependence of the reaction rate is typical for systems driven by thermal fluctuations represented in the form of the Gaussian white noise.
The  escape scenarios driven by non-Gaussian L\'evy noises differ significantly from those induced by thermal fluctuations, in the sense that for
 weak noises the  escape events are performed in single long jumps.
Consequently, the reaction rate is not sensitive to the barrier height but to the barrier width, i.e. $k\propto \delta^{-\alpha}$.
Despite the fact that this approximation is derived in the weak noise limit, it also works pretty well for finite noise strengths.
In a combined action of L\'evy and Gaussian noises one observes competition between L\'evy noise induced long jumps
and contributions of short-length displacements secured by Gaussian part of fluctuations.
As a result, trajectories surmounting the potential start to penetrate the barrier and the escape rate becomes sensitive to the barrier height.
The very same behavior is observed for the noise induced escape from finite intervals where addition of noise with lighter tails increases probability of approaching absorbing boundaries because the likelihood of approaching absorbing boundaries is controlled by the central part of the jump length distribution which is amplified by the additional noise source.

Divergent moments of L\'evy statistics and L\'evy motion seem to stay in conflict with energetic and thermodynamics of the stochastic differential equation of the Langevin type \cite{kusmierz2016breaking,kusmierz2018,kusmierz2014,bier2018}. Yet, accumulating evidence shows that Markovian L\'evy flights (LFs) with distribution of jumps emerging from the generalized version of the central limit theorem are well suited representations of complex phenomena, to name just a few recent applications of  LFs in description of mental searches \cite{baron2013}, analysis of free neutron output in a fusion experiment with a deuteron plasma \cite{ebeling2010convoluted}, investigations of gene-regulatory networks \cite{chen2019} or examination of self-regulatory motion of insects \cite{alsg2018}. 

Long displacements of walkers in fractional dynamics on networks have been shown to improve efficiency to reach any node of the network by inducing small world properties \cite{Mateos2014}, independently of the network structure. This observation is crucial in developing algorithms for optimization based on L\'evy flights techniques. A similar statement can be drawn from the data analysis of option markets which indicate that dispersal of asset prices in actively traded markets is influenced by L\'evy flights or tempered L\'evy flights \cite{Karlova,stanley1986}. Also here, the LFs driven Langevin equation seems to be a proper model of studies, despite infinite variance of fluctuations. The environments powered by L\'evy noise can be natural sources of epicatalytic reactions \cite{bier2018}: whereas in a common catalysis the establishment of equilibrium is speed up by lowering the barrier between two states, in epicatalysis the effect can be achieved by altering the steady state distribution alike to our analysis in Section~\ref{sec:models}. Since also description of various  critical phenomena requires non-local interactions in space (and time) -- it seems plausible to further carefully explore pros and cons of using LF models in realistic applications.


\begin{acknowledgments}
This project was supported by the National Science Center (Poland) grant 2018/31/N/ST2/00598.
This research was supported in part by PL-Grid Infrastructure.
Computer simulations have been performed at the Academic
Computer Center Cyfronet, AGH University of Science and Technology (Krak\'ow, Poland)
under CPU grant DynStoch.

\end{acknowledgments}

%
%
%
\appendix
\section{Mean escape time\label{app:mfpt}}
The two state approximation along with the assumption that escape is performed via the single long jump can be used to calculate the mean first passage time of a free particle from a bounded domain.
For the system described by the Eq.~(\ref{eq:langevin}) the escape takes place under the condition
\begin{equation}
\sigma \xi \Delta t^{1/\alpha}    \geqslant \delta
\end{equation}
leading to
\begin{equation}
 \xi     \geqslant \xi_0  = \frac{\delta}{\sigma} \Delta t^{-1/\alpha}.
\end{equation}
For the $\alpha$ stable density the probability of performing jump longer than $\xi_0$ is $p=P(\xi \geqslant \xi_0) = \xi_0^{-\alpha}$.
Therefore, we obtain the estimation
\begin{equation}
p=P(\xi \geqslant \xi_0)=\frac{\delta^{-\alpha}}{\sigma^{-\alpha}}     \Delta t.
\end{equation}
In order to calculate the mean first passage time, it is necessary to calculate the average number of jumps needed to escape for the first time.
The number of jumps $k$ required to escape for the first time follows the geometric distribution
\begin{equation}
    p_k=(1-p)^{k-1}p,
\end{equation}
because the escape is performed after $(k-1)$ unsuccessful trails.
The mean number of jumps is
\begin{equation}
    \langle k \rangle = \sum_{k=1}^{\infty} p_k k = \frac{1}{p}.
\end{equation}
Since, jumps are performed every $\Delta t$ the MFPT $\langle \tau \rangle$ is
\begin{equation}
    \langle \tau \rangle = \Delta t \langle k \rangle = \frac{\Delta t}{p} = \frac{\delta^\alpha}{\sigma^\alpha}.
    \label{eq:mfpt-app}
\end{equation}
Alternatively, Eq.~(\ref{eq:mfpt-app}) can be derived by investigating scaling of $\langle x^2 \rangle$ with the increasing number of jumps, see Refs.~\onlinecite{szczepaniec2015escape,bouchaud1990}.
Formula (\ref{eq:mfpt}) resembles the general formula for the MFPT \cite{blumenthal1961, getoor1961, kac1950distribution, widom1961stable, kesten1961random} for a particle starting in the middle of the interval of half-width $\delta$ subject to the action of L\'evy noise
\begin{equation}
    \langle \tau \rangle=\frac{\delta^\alpha}{\Gamma(1+\alpha) \sigma^\alpha}.
    \label{eq:mftp-getoor}
\end{equation}

The considerations leading to Eq.~(\ref{eq:mfpt}) do not take into account the process of surmounting the potential barrier. 
Consequently, the escape from the potential well should be not faster than the constructed estimate, see Eqs.~(\ref{eq:mfpt-app}) and (\ref{eq:mftp-getoor}).

 
 
\section*{references} 
 
\def\url#1{}

\end{document}